\documentclass[a4paper,11pt]{article}
\pdfoutput=1 

\usepackage{jheppub} 

\usepackage[T1]{fontenc} 
\usepackage[english]{babel}

\usepackage{graphicx}
\usepackage{amsmath,amsthm,amssymb}
\usepackage{mathrsfs,bbm}
\usepackage{array,color}
\usepackage{slashed}
\usepackage{multirow}
\usepackage{tensor}

\usepackage[load-configurations = abbreviations]{siunitx}

\usepackage{caption}
\usepackage{subcaption}

\newcommand{\PreserveBackslash}[1]{\let\temp=\\#1\let\\=\temp}
\newcolumntype{C}[1]{>{\PreserveBackslash\centering}p{#1}}

\DeclareMathSymbol{\shortminus}{\mathbin}{AMSa}{"39}

\newcommand*{\eweakgroup}{\mbox{$SU(2)_L \times U(1)_Y$} }
\newcommand*{\emgroup}{\mbox{$U(1)_{em}$} }

\newcommand*{\unitmatrix}{\mathbbm{1}}

\newcommand*{\abs}[1]{\left\lvert {#1} \right\rvert} 
\newcommand*{\twomat}[1]{\underline{#1}}             
\newcommand*{\tvec}[1]{\boldsymbol{#1}}              
\newcommand*{\im}[1]{\text{Im} {#1}}                        
\newcommand*{\re}[1]{\text{Re} {#1}}                        

\newcommand*{\trans}{\mathrm{T}}                     

\DeclareMathOperator{\diag}{diag}		
\DeclareMathOperator{\tr}{tr}		
\DeclareMathOperator{\rank}{rank}		

\newcommand{\hide}[1]{}

\title{Multiple point principle in the general Two-Higgs-Doublet model}

\author[a]{Markos Maniatis,}
\author[b]{Lohan Sartore,}
\author[b]{Ingo Schienbein}

\affiliation[a]{Centro de Ciencias Exactas and Departamento de Ciencias Basicas,  Universidad del B\'io-B\'io,  Casilla  447,  Chill\'{a}n,  Chile}
\affiliation[b]{Laboratoire de Physique Subatomique et de Cosmologie, Universit\'e Grenoble-Alpes, CNRS/IN2P3, 53 Avenue des Martyrs, 38026 Grenoble, France}

\emailAdd{maniatis8@gmail.com}
\emailAdd{sartore@lpsc.in2p3.fr}
\emailAdd{schien@lpsc.in2p3.fr}

\abstract{Based on the Multiple Point Principle, the Higgs boson mass has been predicted to be 
$135 \pm 9 \text{ GeV}$ - more than two decades ago. 
We study the Multiple Point Principle and its prospects with respect to the Two-Higgs-Doublet model (THDM).
Applying the bilinear formalism we  show that concise conditions can be given with 
a classification of different kinds of realizations of this principle. We recover cases discussed in the literature but identify also different realizations of the Multiple Point Principle.}

\begin{document} 
\maketitle
\flushbottom

\section{Introduction}
\label{sec:intro}

\subsection{The Multiple Point Principle}
\label{sec:MPP}

The origin of the values for the 19 parameters of the Standard Model is unknown. These parameters are the three lepton masses, six quark masses,
the three coupling strengths, the QCD vacuum angle, the CKM mixing angles along with a phase, 
the electroweak vacuum-expectation value and the Higgs-boson mass.
As an underlying mechanism to generate parameter values, the so-called Mulitple Point Principle (MPP) ~\cite{Bennett:1993pj,Bennett:1996hx,Bennett:1996vy} has been proposed.
This principle may explain how parameters of the Standard Model, like the masses of the Higgs boson and the top quark, come about.
The basic idea of this principle may be illustrated by the triple point of water (as pointed out for 
instance in~\cite{Bennett:2003yr}): At the triple point of water we observe
 three coexisting phases, that is, its solid, liquid, and gaseous form.  
The triple point occurs at specific, that is, fine-tuned values for
the intensive parameters temperature and pressure of about 273.16 K and 0.612 kPa. 
Since the transitions between the three phases are all of first order,
energy and volume can be varied in a certain range without changing neither
temperature nor pressure of the triple point. 
In an analogous way coexisting phases of the Higgs potential may determine
the fine-tuned values of the masses of the bosons and fermions.
Besides one minimum at the electroweak scale of  ${\cal O}(100) \text{ GeV}$ there
should be a degenerate minimum or degenerate minima at a scale $\Lambda$ far above the electroweak scale up to the Planck scale. 
Based on this principle
the Higgs boson mass has been predicted more than 20 years ago to be 
$135 \pm 9 \text{ GeV}$~\cite{Froggatt:1995rt}!
A more refined analysis yielded a mass $129.4 \pm 2$ GeV \cite{Degrassi:2012ry}
to be compared to the observed mass $125.10 \pm 0.14$ GeV \cite{Tanabashi:2018oca} of the Higgs boson
which was discovered by the CMS and ATLAS collaborations in 2012 \cite{Chatrchyan:2012xdj,Aad:2012tfa}.

Let us briefly sketch the derivation of this remarkable result following closely~\cite{Froggatt:1995rt}:
The MPP states that
\begin{enumerate}
\item The Higgs-boson doublet $\varphi$ has at least two coexisting vacua 1 and 2: with the same potential value, that is,
\begin{equation} \label{Cond1}
V_{\text{eff}}(\langle \varphi \rangle_1 )= V_{\text{eff}}(\langle \varphi \rangle_2 ).
\end{equation}
\item The additional minimum or minima should appear at the high scale $\Lambda$ with $100 \text{ GeV}\ll \Lambda <  M_{\text{Planck}}$, 
\begin{equation} \label{Cond2}
\langle \varphi \rangle_2 = {\cal O}(\Lambda).
\end{equation}
\end{enumerate}
In the Standard Model with only one Higgs-boson doublet, the
effective gauge-invariant potential is written, where we define  $\phi = \sqrt{ \varphi^\dagger \varphi}$,
\begin{equation}
V_{\text{eff}} = \mu^2(\phi) \phi^2 + \frac{\lambda(\phi)}{8} \phi^4 \;,
\end{equation}
and where the dependence of the parameters on the scale is written explicitly.
Close to the second vacuum the quartic term is dominant,
\begin{equation}
V_{\text{eff}} \approx \frac{\lambda(\phi)}{8} \phi^4 \;.
\end{equation}
The two conditions (1) and (2) above then give, using the fact that 
the degeneracy of the potential values requires that $\lambda(\Lambda) \approx 0$,
\begin{equation}
\frac{d V_{\text{eff}} }{d \phi} \bigg |_{\text{vac 2}} 
= 0 
 = \frac{1}{2} \lambda (\Lambda) \Lambda^3
 + \frac{1}{8} \frac{d \lambda(\phi)}{d \phi}  \bigg |_{\text{vac 2}}  \Lambda^4 
 = \frac{1}{8} \frac{d \lambda(\phi)}{d \phi}  \bigg |_{\text{vac 2}}  \Lambda^4 
 = \frac{1}{8} \beta_\lambda(\Lambda)  \Lambda^3\;.
\end{equation}
This means that in addition to the quartic parameter $\lambda$ also
its $\beta_\lambda$ function has to vanish at the scale $\Lambda$. 
The $\beta_\lambda$ function depends in the following way on the Higgs-boson field $\phi$:
\begin{equation}
\beta_\lambda (\phi) = \frac{d \lambda(\phi)}{d \ln (\phi)}
=\phi \frac{d \lambda(\phi)}{d (\phi)}
 =\beta_\lambda ( \lambda (\phi), g_t(\phi), g_1(\phi), g_2(\phi), g_3(\phi) )
\end{equation}
with $g_t(\phi)$ the top-Yukawa coupling and $g_{1/2/3}(\phi)$ the scale-dependent gauge couplings. From
the explicit form of the $\beta_\lambda$ function in the Standard Model, Nielsen and Froggatt evaluate the renormalization group equation numerically, using two loop beta functions and plot $\lambda(\phi)$. 
The evolution depends on the masses of the top-quark and the Higgs boson.
Requiring a vanishing quartic parameter $\lambda(\phi)$  as well as a vanishing $\beta_\lambda$ function  at the 
high scale $\Lambda$, the masses of the top quark and the Higgs boson are predicted.\\

This prediction for two of the Standard Model parameters is clearly remarkable and raises the question whether the particular form of the Higgs effective potential 
at the high scale is just an accident or whether there is a deeper law of nature represented by the MPP. 
For this reason, it is worthwhile to have a closer look how the MPP has been motivated in the literature. 
Originally, the MPP was justified using thermodynamical arguments \cite{Bennett:1993pj,Bennett:1996hx,Bennett:1996vy} considering a micro-canonical
ensemble in which an extensive quantity (energy, volume, number of particles) is fixed to a given value. It is then argued that this leads to the MPP.
The system is described by a set of intensive quantitiies (temperature, pressure, chemical potential) and it is assumed that more than one
phase exists with a strong (first order) phase transition between different phases.

It is then argued that in such a system with a fixed extensive variable
the probability is high that intensive quantities (temperature, pressure, chemical potential) take on critical values corresponding to a state with two or more coexisting phases.

Motivated by this result, the question arises  what are the consequences of the MPP in the Two-Higgs-Doublet Model (THDM)?
The original motivation by T. D.~Lee~\cite{Lee:1973iz} to study the two-Higgs-doublet extension
has been to have another source for CP violation -- one of the shortcomings of the Standard Model, where violation of CP only arises from the CKM matrix (and the PMNS matrix) and is too small to explain the observed baryon asymmetry dynamically. Another motivation has been given by supersymmetric models which require to have more than one Higgs-boson doublet in order to give masses to up- and down type fermions. A more pragmatic reason is that there is nothing which prevents the introduction of more copies of Higgs-boson doublets. 
In particular, the $\rho$ parameter, relating the masses of the electroweak gauge bosons with the weak mixing angle (see~\cite{Bernreuther:1998rx} for details) is measured close to one in agreement with the Standard Model.
The $\rho$ parameter is known to keep unchanged at tree level with respect to additional copies of Higgs-boson doublets.
Eventually, let us mention that compared to the two real parameters of the Higgs potential of the Standard Model, the potential of the THDM has a much richer structure allowing for different phases.
For a review of the THDM we refer to~\cite{Branco:2011iw}.

The most general gauge-invariant potential with two Higgs-boson doublets
\begin{equation} \label{doublets}
\varphi_1 = \begin{pmatrix} \varphi_1^{(+)}\\ \varphi_1^{(0)} \end{pmatrix},
\qquad
\varphi_2 = \begin{pmatrix} \varphi_2^{(+)}\\ \varphi_2^{(0)} \end{pmatrix},
\end{equation}
in the convention with both Higgs doublets carrying hypercharge $y=+1/2$, 
reads~\cite{Gunion:1989we}
\begin{equation}
\label{Vconv}
\begin{split}
V^{\text{THDM}} (\varphi_1, \varphi_2) =~& 
m_{11}^2 (\varphi_1^\dagger \varphi_1) +
m_{22}^2 (\varphi_2^\dagger \varphi_2) -
m_{12}^2 (\varphi_1^\dagger \varphi_2) -
(m_{12}^2)^* (\varphi_2^\dagger \varphi_1)\\
& +\frac{1}{2} \lambda_1 (\varphi_1^\dagger \varphi_1)^2 
+ \frac{1}{2} \lambda_2 (\varphi_2^\dagger \varphi_2)^2 
+ \lambda_3 (\varphi_1^\dagger \varphi_1)(\varphi_2^\dagger \varphi_2) \\ 
&+ \lambda_4 (\varphi_1^\dagger \varphi_2)(\varphi_2^\dagger \varphi_1)
+ \frac{1}{2} [\lambda_5 (\varphi_1^\dagger \varphi_2)^2 + \lambda_5^* 
(\varphi_2^\dagger \varphi_1)^2] \\ 
&+ [\lambda_6 (\varphi_1^\dagger \varphi_2) + \lambda_6^* 
(\varphi_2^\dagger \varphi_1)] (\varphi_1^\dagger \varphi_1) + [\lambda_7 (\varphi_1^\dagger 
\varphi_2) + \lambda_7^* (\varphi_2^\dagger \varphi_1)] (\varphi_2^\dagger \varphi_2).
\end{split}
\end{equation}
The parameters $m_{12}^2$, $\lambda_{5/6/7}$ are complex, whereas all other parameters have to be real 
in order to yield a real potential. Therefore we count in total 14 real parameters in contrast to two real parameters of the Standard Model.\\

In the work~\cite{Froggatt:2004st} the MPP has been studied with respect to the general THDM. 
The argumentation in this work has been developed in the following way:
Supposing that the potential
has a second minimum at a high scale $\Lambda$, after an appropriate 
\eweakgroup transformation the vacuum expectation values of the two Higgs-boson doublets are parametrized as \cite{Froggatt:2004st}
\begin{equation} \label{Holgpara}
\langle \varphi_1 \rangle = \phi_1 \begin{pmatrix} 0 \\ 1 \end{pmatrix}, 
\qquad
\langle \varphi_2 \rangle = \phi_2 \begin{pmatrix} \sin(\theta) \\ \cos (\theta) e^{i \omega} \end{pmatrix},
\end{equation}
where $\Lambda^2= \phi_1^2+\phi_2^2$.
Then the conditions are studied for the potential \eqref{Vconv} and its derivatives with respect to $\phi_{1/2}$  to be independent of the phase $\omega$ at the scale $\Lambda$. This leads to conditions for the quartic couplings as well as their derivatives, that is, the $\beta$ functions, evaluated at the scale $\Lambda$.
In particular it is shown that these conditions originating from the MPP yield a CP conserving potential, obeying
in addition a softly broken $Z_2$ symmetry giving an argument for the absence of flavor-changing neutral currents.
Therefore, they continue their analysis in the framework of models with natural flavor conservation, e.g.\ the THDM type II.

{In \cite{McDowall:2018ulq} a detailed phenomenological study of the MPP is carried out, starting from the results of \cite{Froggatt:2004st}, and applying them to the THDM type II as well as the Inert Doublet Model. It is found that in both cases, the MPP is incompatible with the requirement of providing simultaneously the experimental value of the top quark mass, electroweak symmetry breaking and stability.
}

Here we show that the study of the MPP in the THDM can be implemented 
concisely in the bilinear formalism~\cite{Nagel:2004sw,Maniatis:2006fs,Nishi:2006tg}.
The advantage of the bilinear formalism is that all unphysical gauge degrees of freedom are eliminated systematically and the potential and all parameters are real. Basis transformations, that is, 
a unitary mixing of the two doublets, for instance, are given by simple rotations.
Similar to the case of the Standard Model, where we have to have a vanishing quartic coupling together with its $\beta$ function at the high scale $\Lambda$, we find conditions among the potential parameters and its derivatives in order to satisfy the MPP.
We present a classification of all possible realizations of the MPP in the THDM.  
The conditions for these classes of realizations are given in a basis-invariant way and can be checked easily for any THDM. 

In order to arrive at the conditions for the MPP we present the $\beta$ functions
of the potential parameters in the bilinear formalism (see also~\cite{Ma:2009ax}).
We demonstrate the conditions of different realizations of the MPP in examples
and we show in particular that the results of \cite{Froggatt:2004st} can be recovered in the presently developed formalism as one possible realization of the MPP.
It should be noted that other MPP solutions have been discussed in \cite{Froggatt:2008am} in the conventional formalism.
Here, we will present a complete classification of the MPP solutions in a transparent way using the bilinear formalism.

\subsection{Brief review of bilinears in the THDM}
\label{sec:bilinears}

Here we briefly review the bilinears in the THDM~\cite{Nagel:2004sw, Nishi:2006tg, Maniatis:2006fs} in order to make this article 
self contained. We will also discuss briefly basis transformations.
Bilinears systematically avoid unphysical gauge degrees of freedom and are defined in the following way: 
All possible gauge-invariant scalar products of the two doublets $\varphi_1$ and $\varphi_2$ which may appear
 in the potential can be arranged in one matrix 
\begin{equation}
\twomat{K} = 
\begin{pmatrix}
 \varphi_1^\dagger \varphi_1 &  \varphi_2^\dagger \varphi_1\\
 \varphi_1^\dagger \varphi_2 &  \varphi_2^\dagger \varphi_2
 \end{pmatrix}.
 \end{equation}
 This  hermitian matrix can be decomposed into a basis of 
 the unit matrix and the Pauli matrices,
 \begin{equation}
 \twomat{K} = \frac{1}{2} \left(
K_0 \unitmatrix_2 + K_a \sigma_a \right), \qquad a=1,2,3,
\end{equation}
with four real coefficients $K_0$, $K_a$, called bilinears.
Building traces on both sides of this equation (also with products of Pauli matrices) 
we get the four real bilinears explicitly,
\begin{align}
\label{eqKphi}
&K_0 = \varphi_1^\dagger \varphi_1 + \varphi_2^\dagger \varphi_2,
&&K_1 = \varphi_1^\dagger \varphi_2 + \varphi_2^\dagger \varphi_1, \nonumber \\
&K_2 = i\varphi_2^\dagger \varphi_1 - i\varphi_1^\dagger \varphi_2, 
&&K_3 = \varphi_1^\dagger \varphi_1 - \varphi_2^\dagger \varphi_2. 
\end{align}
The matrix $\twomat{K}$ is positive semi-definite. 
From $K_0 = \tr (\twomat{K})$  and $\det(\twomat{K}) = \tfrac{1}{4} (K_0^2 -K_a K_a)$
we get 
\begin{equation} \label{Kdom}
K_0 \ge 0, \qquad K_0^2 - K_a K_a \ge 0.
\end{equation}
As has been shown in~\cite{Maniatis:2006fs} there is a one-to-one correspondence between the original
doublet fields and the bilinears apart from unphysical gauge-degrees of freedom. 
In terms of bilinears
we can write any THDM (a constant term can always be dropped),
\begin{equation}
\label{pot}
V^{\text{THDM}} (K_0, K_a)=
 \xi_0 K_0 + \xi_a K_a + \eta_{00} K_0^2 + 2 K_0 \eta_a K_a + K_a E_{ab} K_b,
\end{equation}
with real parameters $\xi_0$, $\xi_a$, $\eta_{00}$, $\eta_a$, $E_{ab} = E_{ba}$, $a,b \in \{1,2,3\}$.
Expressed in terms of the conventional parameters~\eqref{Vconv}, these new parameters read
\begin{align}
\xi_0 &= \frac{1}{2}\left(m_{11}^2+m_{22}^2\right) ,
\quad
\tvec{\xi} = (\xi_\alpha)=\frac{1}{2}
\begin{pmatrix}
- 2 \re(m_{12}^2), &
 2 \im(m_{12}^2), &
 m_{11}^2-m_{22}^2
\end{pmatrix}^\trans,
\label{eq:para1}
\\
\eta_{00} &= \frac{1}{8}
(\lambda_1 + \lambda_2) + \frac{1}{4} \lambda_3  ,
\quad
\tvec{\eta} = (\eta_{a})=\frac{1}{4}
\begin{pmatrix}
\re(\lambda_6+\lambda_7), & 
-\im(\lambda_6+\lambda_7), & 
\frac{1}{2}(\lambda_1 - \lambda_2)
\end{pmatrix}^\trans, 
\label{eq:para2}
\\
E &= (E_{ab})= \frac{1}{4}
\begin{pmatrix}
\lambda_4 + \re(\lambda_5) & 
-\im(\lambda_5) & \re(\lambda_6-\lambda_7) 
\\ 
-\im(\lambda_5) & \lambda_4 - \re(\lambda_5) & 
-\im(\lambda_6-\lambda_7) \\ 
\re(\lambda_6-\lambda_7) & 
-\im(\lambda_6 -\lambda_7) & 
\frac{1}{2}(\lambda_1 + \lambda_2) - \lambda_3
\end{pmatrix}.
\label{eq:para3}
\end{align}

A unitary basis transformations of the doublets,
\begin{equation}
\varphi'_i = U_{ij} \varphi_j, \quad \text{with } U = (U_{ij}),  \quad U^\dagger U = \unitmatrix_2\, ,
\end{equation}
corresponds to a transformation of the bilinears,
\begin{equation} \label{b1}
K_0' = K_0, \quad K_a' = R_{ab}(U) K_b \;,
\end{equation}
with $R_{ab}(U)$ defined by
\begin{equation}
U^\dagger \sigma^a U = R_{ab}(U) \sigma^b \;.
\end{equation}
It follows that $R(U) \in SO(3)$, that is, $R(U)$ is a proper rotation in three dimensions. 
We see that the potential~\eqref{pot} stays invariant under a 
change of basis of the bilinears \eqref{b1}
if we simultaneously transform the parameters~\cite{Maniatis:2006fs}
\begin{equation} \label{b2}
\xi'_0 = \xi_0, \quad
\xi_a' = R_{ab} \xi_b, \quad
\eta_{00}' = \eta_{00}, \quad
\eta_{a}' = R_{ab} \eta_{b}, \quad
E_{cd}' = R_{ca}  E_{ab} R_{bd}^\trans \, .
\end{equation}
Note that by a change of basis we 
can always diagonalize the real symmetric matrix~$E$. 

{As an illustration, the following parametrization for a unitary transformation~\cite{Maniatis:2006fs}}
\begin{align}\label{EWbasisChange}
    \begin{pmatrix} \varphi_1' \\ \varphi_2' \end{pmatrix} &= 
    \begin{pmatrix} \cos({\beta}) && \sin({\beta})\,e^{- i \zeta} \\ -\sin({\beta})\, e^{i \zeta} && \cos({\beta}) \end{pmatrix}  
    \begin{pmatrix} \varphi_1 \\ \varphi_2 \end{pmatrix} \equiv U \begin{pmatrix} \varphi_1 \\ \varphi_2 \end{pmatrix}
\end{align}
{corresponds, in terms of bilinears, to the rotation matrix}
\begin{equation} \label{unitR}
R(U) = \begin{pmatrix}
 \cos ^2(\beta )-\sin ^2(\beta ) \cos (2 \zeta ) & -\sin ^2(\beta ) \sin (2 \zeta ) & -\sin (2 \beta ) \cos (\zeta ) \\
 -\sin ^2(\beta ) \sin (2 \zeta ) & \sin ^2(\beta ) \cos (2 \zeta )+\cos ^2(\beta ) & -\sin (2 \beta ) \sin (\zeta ) \\
 \sin (2 \beta ) \cos (\zeta ) & \sin (2 \beta ) \sin (\zeta ) & \cos (2 \beta )
\end{pmatrix}
\end{equation}
{
and is useful to relate a given general basis to the so-called Higgs basis (see appendix~\ref{EWbreak}) if the doublets acquire non-zero 
vacuum-expectation values. In this case, the angle $\beta$ fulfils $|v_1^0| \sin{\beta} = |v_2^0| \cos{\beta}$ (or $\tan{\beta} = |v_2^0|/|v_1^0|$). \\
}

Let us also briefly recall that (standard) CP transformations, that is, $\varphi_i \to \varphi_i^*$, $i=1,2$, have a simple geometric picture in terms of bilinears~\cite{Maniatis:2007vn}. With view on~\eqref{eqKphi} we see that a (standard) CP transformation corresponds to $K_2 \to - K_2$ keeping all other bilinears invariant in addition to the
parity transformation which flips the sign of the arguments not written explicitly.
Now let us assume for simplicity, that by a change of basis, the parameter matrix $E$ is diagonal. 
For the general case of arbitrary matrices $E$ we refer to~\cite{Maniatis:2007vn}. 
With $E$ diagonal we see that
the potential~\eqref{pot}  is invariant under the (standard) 
CP transformation if the parameters $\xi_2$ and  $\eta_2$ vanish. 
Eventually we note that by a basis change \eqref{b1}, \eqref{b2} this is equivalent to any commonly vanishing
entries of the parameter vectors  $\tvec{\xi}=(\xi_1, \xi_2, \xi_3)^\trans$ and 
$\tvec{\eta}=(\eta_1, \eta_2, \eta_3)^\trans$ at the same position.

Let us also prepare the analysis of the THDM for the case of large $K_0$.
First we note that $K_0 \ge 0$ and for $K_0=0$ the potential is trivially vanishing.
We define for $K_0 > 0$~\cite{Maniatis:2006fs}
\begin{equation}
\label{eq-ksde}
k_a = \frac{K_a}{K_0}, \quad a \in \{1,2,3\} \quad \text{and} \quad 
\tvec{k} = 
\begin{pmatrix} k_1, & k_2, & k_3 \end{pmatrix}^\trans \;.
\end{equation}
With \eqref{eq-ksde}
we can write the potential~\eqref{pot}
in the form
\begin{equation}
\label{eq-vk}
V^{\text{THDM}} = K_0\, J_2(\tvec{k}) + K_0^2\, J_4(\tvec{k})
\end{equation}
with the functions
\begin{equation}
\label{J2J4}
J_2(\tvec{k}) = \xi_0 + \tvec{\xi}^\trans \tvec{k},\qquad
J_4(\tvec{k}) = \eta_{00} 
  + 2 \tvec{\eta}^\trans \tvec{k} + \tvec{k}^\trans E \tvec{k},
\end{equation}
defined 
on the compact domain, as follows from~\eqref{Kdom},
\begin{equation}
|\tvec{k}| \leq 1 \;.
\end{equation}

In appendix \ref{EWbreak} we recap some parts of
 electroweak symmetry breaking in the THDM in terms of bilinears.


\section{Classification of the vacua}
\label{class}
We now apply the MPP to the THDM potential, that is, we study 
 the two conditions (1) and (2) as mentioned in section~\ref{sec:MPP} for the case of the THDM.
Analogously to the Standard Model case we consider the parameters as scale dependent. 
In the bilinear formalism, advantageous in the 
description of the THDM potential, large field configurations correspond to a
 large bilinear
$K_0$ which itself is bilinear in the Higgs-doublet fields, see \eqref{eqKphi}.
Therefore the 
bilinears depend quadratically on the mass scale $\Lambda$.
The THDM potential is 
considered as an effective parametrization $V_{\text{eff}}^{\text{THDM}}$ of the form \eqref{eq-vk}, \eqref{J2J4}.
Higher-dimensional order operators are neglected since 
we consider a  
scale $\Lambda$ much larger than the electroweak scale but also sufficiently below
the Planck mass.
For large fields close to the high scale the quartic terms are dominant,
so we have
\begin{equation}
\label{Veff}
V_{\text{eff}}^{\text{THDM}} \approx K_0^2\, J_4, \quad \text{ for large $K_0$}
\end{equation}
with 
\begin{equation}
J_4 = \eta_{00} (K_0)
  + 2 \eta_a (K_0) k_a + k_a E_{ab}(K_0) k_b,
\end{equation}
where we write explicitly the dependence of the parameters on the
scale $K_0$. In appendix~\ref{appendixJ4} we briefly discuss the condition of 
a vanishing function $J_4$ at the high scale.
 Note that the bilinear dimensionless field $\tvec{k}$ is 
defined on the domain $|\tvec{k}| \le 1$. 
With respect to the MPP we are looking for a potential
 which has a second minimum at the high scale, that is, 
 $\langle K_0 \rangle_2 = \Lambda^2$ at a 
corresponding ``direction'' of the second minimum, see \eqref{eq-ksde}-\eqref{J2J4},
\begin{equation}
\langle \tvec{k} \rangle_2 \;.
\end{equation}

In order to have a degenerate vacuum at the high scale with the same
potential value we find with respect to \eqref{Veff} for large $K_0$ the condition
\begin{equation}   \label{cond1THDM}
J_4 ( K_0 = \Lambda^2, \langle \tvec{k} \rangle_2 ) = 
\eta_{00} (\Lambda^2)
  + 2 \eta_a (\Lambda^2) \langle k_a \rangle 
  + \langle k_a \rangle E_{ab}(\Lambda^2) \langle k_b \rangle
= 0 \;.
\end{equation}
As a function of $K_0$, $J_4( K_0, \langle \tvec{k}\rangle_2 )$
should provide a minimum with $J_4 = 0$. 
This requires also
\begin{equation}
\left .\frac{ \partial V_{\text{eff}}^{\text{THDM}} } { \partial K_0} \right |_{ \text{vac 2}}
= 0 
= 
 2 \Lambda^2 J_4 \bigg |_{ \text{vac 2}}
+
\Lambda^4
 \frac{ \partial J_4 } { \partial K_0} \bigg |_{ \text{vac 2}}
=
\Lambda^4 
\left . \frac{ \partial J_4 } { \partial K_0} \right |_{ \text{vac 2}} .
\end{equation}
With 
\begin{equation}
2 K_0
\frac{ \partial J_4 } { \partial K_0}
= 
2 K_0
\left(
\frac{ \partial \eta_{00}(K_0) } { \partial K_0}
+
2 \frac{ \partial \eta_a (K_0) } { \partial K_0} 
k_a
+
k_a
\frac{ \partial E_{ab}(K_0) } { \partial K_0} 
k_b
\right)
=
\beta_{\eta_{00}} + 2 \beta_{\eta_{a}} 
k_a
+ 
k_a
 \beta_{E_{ab}} 
k_b
 \, ,
\end{equation}
we find the condition for the beta functions at the Planck scale:
\begin{equation} \label{cond2THDM}
\beta_{\eta_{00}} (\Lambda^2) + 2 \beta_{\eta_{a}}(\Lambda^2)   \langle {k}_a  \rangle_2 + 
\langle {k}_a \rangle_2 \beta_{E_{ab}}(\Lambda^2)  \langle {k}_b \rangle_2 = 0.
\end{equation}
For the THDM the conditions \eqref{cond1THDM} and \eqref{cond2THDM}  
replace the ones for the SM. 

\subsection{Stationary points at the high scale $\Lambda$}
\label{statPoints}

So far we have found that a minimum $\langle \tvec{k} \rangle_2$ should satisfy
the conditions \eqref{cond1THDM} and \eqref{cond2THDM}.
We now study the stationarity structure of the dominant quartic terms.
This study is quite analogous to the stability study in~\cite{Maniatis:2006fs}. 
We have $|\tvec{k} | \le 1$ and we  consider the cases
 $|\tvec{k} | < 1$ and  $|\tvec{k} | = 1$ separately.

For $|\tvec{k} | < 1$, stationarity of the potential requires,
expressed in terms of $J_4$ (since $K_0 >0$),
\begin{equation}
\nabla_{\tvec{k}} J_4 (\tvec{k})
=
 2 \tvec{\eta}^\trans + 2 \tvec{k}^\trans E  = 0, 
\end{equation}
that is, since $E$ is symmetric
\begin{equation} \label{con}
\tvec{\eta} + E  \tvec{k} =0\;.
\end{equation}
Note that we do not write explicitly the scale dependence of the parameters which 
implicitly is given by $\Lambda^2$.
With the condition \eqref{con} for a vanishing gradient we can 
in the case $|\tvec{k} | < 1$ write the condition for vanishing $J_4$ \eqref{cond1THDM} 
now in the form
\begin{equation} \label{cond12}
\tvec{\eta}^\trans \langle \tvec{k} \rangle_2 = - \eta_{00} \;.
\end{equation}

For $\det (E) \neq 0$ the regular solution of \eqref{con} is
\begin{equation} \label{solreg}
\langle \tvec{k} \rangle_2 = - E^{-1} \tvec{\eta}
\end{equation}
or we have for $\det (E) = 0$ exceptional solutions. 
We check that for the regular solutions we have with $1-\tvec{k}^\trans \tvec{k} > 1$ indeed
\begin{equation} \label{J4reg}
J_4 (\langle \tvec{k} \rangle_2)
=
 \eta_{00} - \tvec{\eta}^\trans E^{-1} \tvec{\eta}  = 0, 
 \quad
 \text{if }
 1 - \tvec{\eta}^\trans E^{-2} \tvec{\eta} >0 .
\end{equation}

For $|\tvec{k} | = 1$ we impose a Lagrange multiplier,
and the stationary solutions follow from 
\begin{equation}
\nabla_{\tvec{k}, u} (J_4 (\tvec{k}) + u (1-\tvec{k}^2) )
= 0 \;,
\end{equation}
that is,
\begin{equation} \label{eqbord}
(E-  \unitmatrix_3 u) \tvec{k} + \tvec{\eta} = 0, \quad  
(1-\tvec{k}^\trans \tvec{k})=0 \;.
\end{equation}
With these conditions for a vanishing gradient we can 
in the case $|\tvec{k} | = 1$ write the condition for a vanishing $J_4$ \eqref{cond1THDM} 
now in the form
\begin{equation} \label{cond12u}
\tvec{\eta}^\trans \langle \tvec{k} \rangle_2 = - \eta_{00} -u \;.
\end{equation}

The regular solution of \eqref{eqbord}, that is, a solution 
with $\det(E- \unitmatrix_3 u) \neq 0$ is
\begin{equation} \label{solbord}
\langle \tvec{k} \rangle_2 = - (E- \unitmatrix_3 u)^{-1} \tvec{\eta}\;,
\end{equation}
where $u$ follows from $(1-\tvec{k}^\trans \tvec{k})=0$ with \eqref{solbord} from
\begin{equation} \label{solu}
 1 - \tvec{\eta}^\trans (E- \unitmatrix_3 u)^{-2} \tvec{\eta} = 0 \;.
\end{equation}
 We check that for the regular solution we have indeed
 \begin{equation}
 J_4 (\langle \tvec{k} \rangle_2)
=
 u + \eta_{00} - \tvec{\eta}^\trans E^{-1} \tvec{\eta}  = 0, 
\end{equation}
with $u$ the solution of \eqref{solu}.
Alternatively, we may have for $\det(E- \unitmatrix_3 u) = 0$
exceptional solutions of \eqref{eqbord}.


Eventually we note that we have to ensure that the 
stationary solutions of \eqref{con}
for $|\tvec{k}| < 1$, respectively, \eqref{eqbord} for $|\tvec{k}| = 1$ are local minima. 
As usual this can be 
done by considering the (bordered) Hessian matrix. Alternatively, in 
case of a stable potential, that is, a potential which is bounded from below, we 
may look for the deepest stationary solution or in the degenerate case, solutions,
which are then of course minima.

\subsection{Classification of the MPP in the THDM}
\label{classification}

Let us now study the vacuum structure with respect to the MPP in detail.
Especially, we derive the conditions to have isolated points, respectively, continuous
stationarity regions, corresponding to the MPP in a weaker or 
a stronger sense. 
First we recall that we can, by a basis change, \eqref{b1}, \eqref{b2}, diagonalize the real symmetric matrix $E$ and therefore we suppose to have
\begin{equation}
E' = \diag (E_{11}', E_{22}', E_{33}' ) .
\end{equation}
We emphasize that $E'$ diagonal is assumed to hold at the scale $\Lambda^2$. This in particular means that in principle the matrix $E'$ may be non-diagonal at a different scale. We discuss the running
of the parameters of the THDM in the next section.

In order to distinguish the parameters in the new basis, where the matrix $E'$ is diagonal, from the
original ones, we denote them with a prime symbol.
In conventional notation the potential with a diagonal matrix $E$ corresponds
to arbitrary parameters with  
$\lambda_6=\lambda_7$ general complex and $\lambda_5$ real.

For $K_0 \neq 0$ the bilinear space is defined on the domain $|\tvec{k}| \le 1$. 
Let us first consider the case $|\tvec{k}| < 1$. 

As pointed out above, for $\det (E) = \det (E') \neq 0$ the regular solution of
the gradient equation \eqref{con}
is a single point \eqref{solreg} and requires that $\eta_{00}$ also satisfies \eqref{cond12},
providing a degenerate value of the potential. 
We emphasize that the condition $\det (E) \neq 0$ as well as the parameter $\eta_{00}$ 
are invariant under a change of basis.

If one of the eigenvalues of $E$ is zero, say in the 
diagonalized matrix $E'$ its upper component, 
we get from \eqref{con} 
\begin{multline}
E'_{11} = 0,\; E'_{22} \neq 0,\; E'_{33} \neq 0  :\\
\quad \text{solution } \langle  \tvec{k} \rangle_2 = 
\begin{pmatrix} x, &  -\frac{\eta_2'}{E_{22}'}, & - \frac{\eta_3'}{E_{33}'} \end{pmatrix}^\trans
\text{ with } 
x^2 < 1 -  \frac{\eta_2'^2}{E_{22}'^2}  -  \frac{\eta_3'^2}{E_{33}'^2}  \text{ for } \eta_1' =0.
\end{multline}
 This is a line segment. 
In the case that $\eta_1'$ together with $E_{11}'$ are vanishing, that is,
 the two zero components of $\tvec{\eta'}$ and $E'$  
 are aligned, 
 we may have a degenerate line of solutions satisfying \eqref{con}. 
For $\eta_1' \neq 0$  there is no solution of \eqref{con}.
 Besides, the
 $\eta_{00}$ parameter has in any case to satisfy \eqref{cond12}.
 We want to derive these conditions in a basis-invariant way. 
Firstly, we remark that one vanishing eigenvalue of the matrix $E$ (note that $E$ is 
the original parameter matrix and not necessarily diagonal) corresponds to 
$\rank(E) = 2$ and this in turn gives the basis-invariant conditions
\begin{equation} \label{Erank2}
 \rank(E) = 2: \qquad  \det(E) = 0 \text{ and } \tr^2(E) - \tr(E^2)  \neq 0 \;.
\end{equation}
Now we can construct the conditions to have one zero in the parameter vector $\tvec{\eta}'$
aligned with $E'$ in a basis-invariant way:
\begin{equation} \label{1align}
(\tvec{\eta} \times (E \tvec{\eta}))^\trans E^2 \tvec{\eta} = 0,
\quad \text{and} \quad
(E^2 \tvec{\eta} \times (E \tvec{\eta}))^\trans (E^2 \tvec{\eta} \times (E \tvec{\eta})) \neq 0 \;.
\end{equation}
We get the statement, that for a matrix $E$ of rank 2, that is, a matrix $E$ fulfilling the conditions \eqref{Erank2},
there is a line of second degenerate vacua possible if in addition the two conditions
\eqref{1align} hold and the basis-invariant parameter 
$\eta_{00}$ satisfies \eqref{cond12}. This corresponds to the MPP in the stronger sense.
 In case that only the rank conditions \eqref{Erank2} hold but not the conditions \eqref{1align} there is no realization 
 of the MPP possible. 

Similarly we can treat the case that two eigenvalues of $E$ vanish. Going to a basis where $E'$ is diagonal, we suppose
that the two upper components of the diagonal matrix $E'$ vanish,  
then we find from \eqref{con} the solutions
\begin{multline} \label{2zero}
E'_{11} =  E'_{22} = 0,\; E'_{33} \neq 0:\\
\text{solution } \langle  \tvec{k} \rangle_2 = 
\begin{pmatrix} x, & y , &  -\frac{\eta_3'}{E_{33}'} \end{pmatrix}^\trans
\text{ with } x^2 + y^2  < 1 -  \frac{\eta_3'^2}{E_{33}'^2},\text{ for }
\eta_1'=\eta_2'=0.
\end{multline}
This solution is a disk. In case 
that not both components of $\tvec{\eta'}$ aligned with
the vanishing diagonal entries of  $E'$ vanish,  there is no solution.
We note that for a solution also the parameter $\eta_{00}$ has to satisfy \eqref{cond12}.

The formulation in a basis-invariant way is as follows:
Two vanishing eigenvalues correspond to a matrix $E$ of $\rank(E) =1$, that is,
\begin{equation} \label{Erank1}
 \rank(E) = 1:\qquad  \det(E) = 0 ,
 \text{ and } \tr^2(E) - \tr(E^2)  = 0 ,
 \text{ and }
\tr(E) \neq 0\;.
\end{equation}
Since two eigenvalues vanish, we can by a basis change always achieve 
that also one of the components of $\tvec{\eta}$ vanish, aligned with one of the vanishing
entries of $E$. This can be written in a basis-invariant way:
\begin{equation} \label{cond2}
\tvec{\eta}^\trans E \tvec{\eta} \neq 0, 
\quad
\text{and}
\quad
(E \tvec{\eta} \times \tvec{\eta})^\trans (E \tvec{\eta} \times \tvec{\eta}) = 0 \;.
\end{equation}
Only in case that in addition to \eqref{Erank1} 
the conditions \eqref{cond2} are satisfied,
we can have the MPP in form of a disk in bilinear space, that is, the MPP in the stronger sense.
We note, that in this case it is required that the parameter 
$\eta_{00}$ satisfies \eqref{cond12}.
Otherwise, if the rank 1 conditions are fulfilled, but \eqref{cond2} do not both hold, there is no solution as a multiple point.

Eventually, we consider the case $E=0$. Note that a vanishing
matrix $E$ does not depend on the chosen basis. Now, we find from \eqref{cond12},
that is, the condition of a vanishing potential at the second minimum,
 that there is for $\tvec{\eta} \neq 0$ no solution with respect to the MPP.
 However, if we have in addition to $E=0$ also $\tvec{\eta} = 0$ and $\eta_{00}=0$ 
we have a sphere of solutions, that is, the MPP realized in the stronger sense.


The case $|\tvec{k}| =1$ can be treated analogously to the previous one. 
We look for solutions of the gradient equation \eqref{eqbord} instead of 
\eqref{con}.
This system of four equations has in general solutions for the indeterminants
$\langle \tvec{k} \rangle_2$ as well as the Lagrange multiplier $u$.
The solutions of \eqref{eqbord} and in particular the degeneracy of the solutions depend on the determinant of the matrix
\begin{equation}
M=E - \unitmatrix_3 u \;.
\end{equation}
Again we get for  $\det{M} \neq 0$
 from \eqref{eqbord} solution points \eqref{solbord}. 
The determinant is of course invariant under basis changes.
 
Let us now turn to the exceptional cases, with at least one  
eigenvalue of the matrix $M$ vanishing.
The argumentation is quite analogously to the previous study where
we have to replace the matrix $E$ by $M$ and have to
take into account the condition $\tvec{k}^\trans \tvec{k} =1$. 

If one of the eigenvalues of $M$ is zero, say, without loss of generality in the 
diagonalized matrix $M'$ its upper component, 
we get from \eqref{con} 
\begin{multline}
M'_{11} = 0,\; M'_{22} \neq 0,\; M'_{33} \neq 0  :\\
\text{solution } \langle  \tvec{k} \rangle_2 = 
\begin{pmatrix} x, &  -\frac{\eta_2'}{M_{22}'}, & - \frac{\eta_3'}{M_{33}'} \end{pmatrix}^\trans
\quad \text{with } 
x^2 = 1 -  \frac{\eta_2'^2}{M_{22}'^2}  -  \frac{\eta_3'^2}{M_{33}'^2},\text{ for }
\eta_1' =0 .
 \end{multline}
 This gives at most two points, supposed that $\eta_1' =0$, and
 otherwise there is no solution. 
 In addition, the $\eta_{00}$ parameter has to satisfy \eqref{cond12u} in order to 
 give a degenerate second vacuum.
 We want to find the conditions independent of the chosen basis. 
One vanishing eigenvalue of the matrix $M$ corresponds to 
$\rank(M) = 2$, hence, basis-invariantly written,
\begin{equation} \label{Erank2u}
 \rank(M) = 2: \qquad  \det(M) = 0 \text{ and } \tr^2(M) - \tr(M^2)  \neq 0 \;.
\end{equation}
The conditions to have one zero in the parameter vector $\tvec{\eta'}$
aligned with the vanishing eigenvalue in $M$ are
\begin{equation} \label{1alignM}
(\tvec{\eta} \times (M \tvec{\eta}))^\trans M^2 \tvec{\eta} = 0,
\quad \text{and} \quad
(M^2 \tvec{\eta} \times (M \tvec{\eta}))^\trans (M^2 \tvec{\eta} \times (M \tvec{\eta})) \neq 0 \;.
\end{equation}
In case there is a solution of  a vacuum with $u$ from \eqref{eqbord} satisfying the conditions \eqref{Erank2u}
and also the conditions \eqref{1alignM},
there are points as a second degenerate vacuum possible
supposed that $\eta_{00}$ satisfies \eqref{cond12u}.

Suppose now that the two components, say, the upper components of the diagonalized matrix $M'$ vanish, then we find from \eqref{eqbord} the solutions
\begin{multline} \label{2zerou}
M'_{11} =  M'_{22} = 0,\; M'_{33} \neq 0:\\
\text{solution } \langle  \tvec{k} \rangle_2 = 
\begin{pmatrix} x, & y , &  -\frac{\eta_3'}{M_{33}'} \end{pmatrix}^\trans
\quad \text{with } x^2 + y^2  = 1 -  \frac{\eta_3'^2}{M_{33}'^2},\text{ for }
\eta_1' = \eta_2' =0.
\end{multline}
Only in case that we have also the two vanishing components of $\tvec{\eta'}$ 
aligned with the vanishing eigenvalues of $M'$, we 
get a circle of degenerate solutions; otherwise there
is no solution.

The formulation in a basis-invariant way is as follows:
Two zero eigenvalues correspond to a matrix $M$ of rank one, that is,
\begin{equation} \label{Erank1u}
 \rank(M) = 1:\qquad  \det(M) = 0 ,
 \text{ and } \tr^2(M) - \tr(M^2)  \neq 0 ,
 \text{ and }
\tr(M) \neq 0\;.
\end{equation}
Since two eigenvalues vanish, we can by a further change of basis always achieve 
that one of the components of $\tvec{\eta}'$ vanishes, aligned with one of the vanishing
entries of $M'$. 
To this purpose we construct two additional conditions, manifestly basis invariant,
\begin{equation} \label{cond2u}
\tvec{\eta}^\trans M \tvec{\eta} \neq 0, 
\quad
\text{and}
\quad
(M \tvec{\eta} \times \tvec{\eta})^\trans (M \tvec{\eta} \times \tvec{\eta}) =0.
\end{equation}
Only in the case that both conditions \eqref{Erank1u} and \eqref{cond2u} hold,
with $\eta_{00}$ satisfying \eqref{cond12u} we can have the MPP realized in form of a 
circle in bilinear space.
In all other cases there is no second vacuum possible.

Eventually, we consider the case $M=0$. Then, we find from \eqref{cond12u},
that is, a vanishing potential at the second minimum,
 that there is for $\tvec{\eta} \neq 0$ no solution with respect to the MPP and
in case of $\tvec{\eta} = 0$ we have a surface of a sphere of solutions, that is, 
the MPP in the stronger sense.
Note that these conditions are already basis invariant. 

%
%

Let us mention that we have seen, that
in addition to a vanishing eigenvalue of the matrix $E$, respectively, 
$M=E - u \unitmatrix_3$, also the corresponding component 
of $\tvec{\eta'}$ has to vanish (in a basis where $E'$, respectively $M'$, is diagonal). 
This in turn means that we do have CP conservation in
this case~\cite{Maniatis:2007vn}.
We thus confirm the result~\cite{Froggatt:2004st}
that the MPP in the THDM in the stronger sense corresponds
to a CP conserving potential.
In the strongest case where the second vacuum is a degenerate sphere we have 
to have
$E = 0$ together with $\tvec{\eta}$ and $\eta_{00}$ vanishing. 
This means that the potential has $J_4 =0$ for all $\tvec{k}$. 
In conventional notation this
gives $\lambda_i = 0$, $i=1,\ldots,7$.

Moreover, let us note the interesting aspect of solutions corresponding 
to $|\tvec{k}| < 1$ (see \cite{Maniatis:2006fs} for details),
which give charge-breaking minima and solutions corresponding to
$|\tvec{k}| = 1$ 
which give  electroweak symmetry breaking \eweakgroup $\to$ \emgroup, 
however, for a second vacuum at
a high vacuum expectation scale $\Lambda$, there is no reason to discard the possibility of charge-breaking minima.

We summarize our findings in table \ref{tabMPP}.

\begin{table}[h!]
\begin{tabular}{lllll}
\hline
\hline
$\det(E) \neq 0$ & & & & point\\
\hline
\multirow{12}{*}{$\det(E) = 0$}
	 & \multirow{6}{*}{$\tr^2(E)-\tr(E^2) \neq 0$}
			& & \multicolumn{2}{l}{\hspace{-24pt}$(\tvec{\eta} \times (E \tvec{\eta}))^\trans E^2 \tvec{\eta} = 0$}
\\
& & & and & line\\
& & & 
\multicolumn{2}{l}{\hspace{-24pt}$(E^2 \tvec{\eta} \times (E \tvec{\eta}))^\trans (E^2 \tvec{\eta} \times (E \tvec{\eta})) \neq 0$}
\\
	 		\cline{4-5}
& & & \multicolumn{2}{l}{\hspace{-24pt}$(\tvec{\eta} \times (E \tvec{\eta}))^\trans E^2 \tvec{\eta} \neq 0$}\\
& & &  or 	 &  no \\
& & & \multicolumn{2}{l}{\hspace{-24pt}$(E^2 \tvec{\eta} \times (E \tvec{\eta}))^\trans (E^2 \tvec{\eta} \times (E \tvec{\eta})) = 0$}	
\\
	 \cline{2-5}
	 &  \multirow{8}{*}{$\tr^2(E)-\tr(E^2) = 0$}
	 	& \multirow{2}{*}{$\tr(E) \neq 0$}
	 		&
	 		$\tvec{\eta}^\trans E \tvec{\eta} \neq 0$\\
& & &	\text{ and } & disk\\
& & & $(E \tvec{\eta} \times \tvec{\eta})^\trans (E \tvec{\eta} \times \tvec{\eta}) = 0$ \\
	 		\cline{4-5}
	 		& & 	& 
 		$\tvec{\eta}^\trans E \tvec{\eta} = 0$ \\
& & &			\text{ or } & no\\
& & &			$(E \tvec{\eta} \times \tvec{\eta})^\trans (E \tvec{\eta} \times \tvec{\eta}) \neq 0$\\ 
	 	\cline{3-5}
		& &  \multirow{2}{*}{$E=0$} & $\tvec{\eta} \neq 0$  & no\\
		\cline{4-5}
 		& &  & $\tvec{\eta} = 0$ & sphere\\
\hline
\hline
$\det(M) \neq 0$ & & & & point\\
\hline
\multirow{5}{*}{$\det(M) = 0$}
	 & \multirow{6}{*}{$\tr^2(M)-\tr(M^2) \neq 0$}
			& & \multicolumn{2}{l}{\hspace{-24pt}$(\tvec{\eta} \times (M \tvec{\eta}))^\trans M^2 \tvec{\eta} = 0$}
\\
& & & and & point\\
& & & 
\multicolumn{2}{l}{\hspace{-24pt}$(M^2 \tvec{\eta} \times (M\tvec{\eta}))^\trans (M^2 \tvec{\eta} \times (M \tvec{\eta})) \neq 0$}
\\
	 		\cline{4-5}
& & & \multicolumn{2}{l}{\hspace{-24pt}$(\tvec{\eta} \times (M \tvec{\eta}))^\trans M^2 \tvec{\eta} \neq 0$}\\
& & &  or 	 &  no \\
& & & \multicolumn{2}{l}{\hspace{-24pt}$(M^2 \tvec{\eta} \times (M \tvec{\eta}))^\trans (M^2 \tvec{\eta} \times (M \tvec{\eta})) = 0$}	\\
	 \cline{2-5}
	 &  \multirow{8}{*}{$\tr^2(M)-\tr(M^2) = 0$}
	 	& \multirow{2}{*}{$\tr(M) \neq 0$}
	 		&
	 		$\tvec{\eta}^\trans M \tvec{\eta} \neq 0$\\
& & &			\text{ and } & circle\\
& & &			$(M \tvec{\eta} \times \tvec{\eta})^\trans (M \tvec{\eta} \times \tvec{\eta}) = 0$\\
	 		\cline{4-5}
	 		& & 	& 
	 		$\tvec{\eta}^\trans M \tvec{\eta} = 0$\\
& & &		 or & no\\
& & &		$(M \tvec{\eta} \times \tvec{\eta})^\trans (M \tvec{\eta} \times \tvec{\eta}) \neq 0$
		  \\
		 	\cline{3-5}
		& &  \multirow{2}{*}{$M=0$} & $\tvec{\eta} \neq 0$  & no\\
		\cline{4-5}
 		& &  & $\tvec{\eta} = 0$ & surface\\
 \hline
 \hline
\end{tabular}
\caption{
\label{tabMPP} 
Classification of possible realizations of the MPP in the THDM. 
The last column gives the kind of realization of the MPP or ``no'' in 
case the MPP is not realized. The upper part gives solutions for 
the case $|\tvec{k}|<1$ and the lower part for the case $|\tvec{k}|=1$.
In all cases where the MPP is realizable, the parameter $\eta_{00}$ has
to fulfill the condition \eqref{cond12}, respectively, \eqref{cond12u}. The solutions
of the vacuum vector $\langle \tvec{k} \rangle$ follow from \eqref{con}, respectively,
the solutions of $\langle \tvec{k} \rangle$ and $u$ from \eqref{eqbord}.
The conditions are given in a basis-invariant way and
are therefore directly applicable to any THDM.
}
\end{table}


\subsection{Constraints from the quantum potential}
\label{quantum}

Thanks to the bilinear formalism, the 1-loop $\beta$-functions of the general THDM can be put in a concise tensor form (see appendix~\ref{rges}), allowing one to perform an analytical study of the renormalization group. In the case where $|\tvec{k}| < 1$, using~\eqref{con} and~\eqref{cond12}, the constraint~\eqref{cond2THDM} can be put into 
the following form:
\begin{align}\label{nullDK0}
    \begin{split}
    0 ={}& 8\left[\left( \tvec{k}^\trans E \tvec{k}\right)^2 - 2\, \tvec{k}^\trans E^2 \tvec{k} + \tr \left( E^2 \right) \right] + g \tvec{k}^\trans \tvec{k} + G\\
    &-\frac{1}{2}\Big\{ \tensor{\left(\sigma_0 + \tvec{k}^\trans \tvec{\sigma}\right)}{^a_b}\tensor{\left(\sigma_0 + \tvec{k}^\trans\tvec{\sigma}\right)}{^c_d} \Big\}\,\tensor{\mathcal{T}}{_a^b_c^d} + \left(1 - \tvec{k}^\trans \tvec{k} \right) \mathcal{T}_{\mathcal{U}\mathcal{D}}\,,
    \end{split}
\end{align}
where we defined for convenience the strictly positive quantities 
\begin{equation}
    g \equiv \frac{9}{20} g_1^2 g_2^2 \quad\text{and}\quad G \equiv \frac{9}{8}\left(\frac{3 g_1^4}{25}+g_2^4\right)
\end{equation}
and where the definition of the $\mathcal{T}$-tensors is given in~\eqref{tTrace}.
If $|\tvec{k}| = 1$, we must use \eqref{eqbord} and~\eqref{cond12u} and the constraint reads 
\begin{align}\label{nullDK0u}
    \begin{split}
    0 ={}& 8\left[ 2 u^2 + \left( \tvec{k}^\trans M \tvec{k}\right)^2 - 2\, \tvec{k}^\trans M^2 \tvec{k} + \mathrm{Tr}\left( M^2 \right) \right] + g + G \\
    & -\frac{1}{2}\Big\{ \tensor{\left(\sigma_0 + \tvec{k}^\trans \tvec{\sigma}\right)}{^a_b}\tensor{\left(\sigma_0 + \tvec{k}^\trans \tvec{\sigma}\right)}{^c_d} \Big\}\,\tensor{\mathcal{T}}{_a^b_c^d} \,.
    \end{split}
\end{align}

Let us first consider a simplified situation where the theory does not contain Yukawa couplings. In that case, eq.~\eqref{nullDK0} becomes
\begin{align}
    0 ={}& 8\left[\left( \tvec{k}^\trans E \tvec{k}\right)^2 - 2\, \tvec{k}^\trans E^2 \tvec{k} + \mathrm{Tr}\left( E^2 \right) \right] + g \tvec{k}^\trans \tvec{k} + G \,,
\end{align}
and has in fact no solutions. This can be seen by working in a basis where $\tvec{k} = (k_1,0,0)^\trans$, and where the term in the brackets now reads
\begin{align}
\begin{split}
    \left( \tvec{k}^\trans E \tvec{k}\right)^2 - 2\, \tvec{k}^\trans E^2 \tvec{k} + \mathrm{Tr}\left( E^2 \right) ={}& k_1^4 E_{11}^2 -2 k_1^2 \left(E_{11}^2+ E_{12}^2+E_{13}^2\right) \\
   &+E_{11}^2+E_{22}^2+E_{33}^2 + 2 (E_{12}^2+ E_{13}^2 + E_{23}^2) \\
   ={}&(1-k_1^2)^2 E_{11}^2 + E_{22}^2 + E_{33}^2 + 2(1-k_1^2)\big[ E_{12}^2 + E_{13}^2 + E_{23}^2 \big]\,.
\end{split}
\end{align}
Clearly, since $\tvec{k}^\trans \tvec{k} <1$ the right-hand side is a positive quantity. This holds in any basis, and implies in turn that 
\begin{align}
    8\left[\left( \tvec{k}^\trans E \tvec{k}\right)^2 - 2\, \tvec{k}^\trans E^2 \tvec{k} + \tr \left( E^2 \right) \right] + g \tvec{k}^\trans \tvec{k} + G > 0\,.
\end{align}

We therefore have proven that in absence of Yukawa couplings, and if $\tvec{k}^\trans \tvec{k} < 1$, eq.~\eqref{cond2THDM} cannot be satisfied, which means that the MPP cannot be realized. Applying the above reasoning to the case $\tvec{k}^\trans \tvec{k}=1$ we get the same conclusion.\\

The situation changes if we consider Yukawa couplings. For the known fermions we will see that only the dominant 
contribution from the top quark allows for a realization of the MPP. Alternatively, lower bounds on the Yukawa couplings
could be given in order to have the MPP realized.

\subsection{Comparison with previous work on the MPP in the THDM}
\label{comparisonPreviousWorks}

As a concrete application of the results derived in sections~\ref{classification} and~\ref{quantum}, let us take the example of a 
THDM type II\footnote{In fact, the following discussion remains valid for any THDM where $\lambda_6 = \lambda_7 = 0$ and where 
none of the fermions couple simultaneously to both doublets.}. 
Thus we will be able to compare our results with the ones from \cite{Froggatt:2004st} and \cite{McDowall:2018ulq}.\\

The scalar potential 
considered in \cite{Froggatt:2004st,McDowall:2018ulq} is the general potential but with the restriction $\lambda_6 = \lambda_7 = 0$. 
With these assumptions the quartic couplings \eqref{eq:para2}, \eqref{eq:para3} simplify as:
\begin{align}
\eta_{00} &= \frac{1}{8}
(\lambda_1 + \lambda_2) + \frac{1}{4} \lambda_3  ,
\quad
\tvec{\eta} = (\eta_{a})=\frac{1}{4}
\begin{pmatrix}
0, & 
0, & 
\frac{1}{2}(\lambda_1 - \lambda_2)
\end{pmatrix}^\trans, 
\\
E &= (E_{ab})= \frac{1}{4}
\begin{pmatrix}
\lambda_4 + \re(\lambda_5) & 
-\im(\lambda_5) & 0
\\ 
-\im(\lambda_5) & \lambda_4 - \re(\lambda_5) & 
0 \\ 
0 & 
0 & 
\frac{1}{2}(\lambda_1 + \lambda_2) - \lambda_3
\end{pmatrix}.
\end{align}


By a basis change we can diagonalize the matrix $E$ without changing $\eta_{00}$, $\tvec{\eta}$. In the new basis the parameter $\lambda_5$ is real,
\begin{align}
E &= (E_{ab})= \frac{1}{4}
\begin{pmatrix}
\lambda_4 + \lambda_5 & 
0 & 0
\\ 
0 & \lambda_4 - \lambda_5 & 
0 \\ 
0 & 
0 & 
\frac{1}{2}(\lambda_1 + \lambda_2) - \lambda_3
\end{pmatrix}.
\end{align}

The most general form for the vacuum expectation values of the doublets may be parametrized as \cite{Froggatt:2004st}
\begin{equation} 
\langle \varphi_1 \rangle = \phi_1 \begin{pmatrix} 0 \\ 1 \end{pmatrix}, 
\qquad
\langle \varphi_2 \rangle = \phi_2 \begin{pmatrix} \sin(\theta) \\ \cos (\theta)\, e^{i \omega} \end{pmatrix},
\end{equation}
where $\Lambda^2= \phi_1^2+\phi_2^2$. In terms of the bilinear fields, this corresponds to 
\begin{align}
	\begin{split}\label{Ks}
		K_0 &= \phi_1^2 + \phi_2^2 = \Lambda^2,\\
		K_1 &= 2 \phi_1 \phi_2  \cos(\theta) \cos(\omega),\\
		K_2 &= 2 \phi_1 \phi_2  \cos(\theta) \sin(\omega),\\
		K_3 &= \phi_1^2 - \phi_2^2\,,
	\end{split}
\end{align}
so there is a direct mapping between $(\phi_1, \phi_2, \theta, \omega)$ and $(K_0, K_1, K_2, K_3)$. Since this will be useful in the following, we note that
\begin{equation}
	\tvec{k}^\trans \tvec{k} 
	= \frac{\tvec{K}^\trans \tvec{K}}{K_0^2} 
	= \frac{1}{\left(\phi_1^2 + \phi_2^2\right)^2}\left( \phi_1^4 + 2 \cos(2\theta) \phi_1^2 \phi_2^2 + \phi_2^4 \right )\,.
\end{equation}
Therefore we find that for $\cos(2\theta) = 1$ or, equivalently, $\cos(\theta) = \pm 1$ we have 
$\tvec{k}^\trans \tvec{k} =1$.\\

We can now begin the comparison between the present work and the previous analysis of the MPP by Froggatt and Nielsen. The study in \cite{Froggatt:2004st} results in two possible vacuum configurations, namely
\begin{subequations}
	\begin{equation}\label{chargeCons}
		\langle \varphi_1 \rangle = \begin{pmatrix} 0 \\  \phi_1 \end{pmatrix}, 
\qquad
\langle \varphi_2 \rangle = \begin{pmatrix} 0 \\  e^{i \omega} \phi_2 \end{pmatrix},
	\end{equation}
	\begin{equation}\label{chargeBr}
		\langle \varphi_1 \rangle =\begin{pmatrix} 0 \\  \phi_1  \end{pmatrix}, 
\qquad
\langle \varphi_2 \rangle = \begin{pmatrix}  \phi_2 \\ 0 \end{pmatrix},
	\end{equation}
\end{subequations}

which respectively correspond to a charge-conserving, CP-violating and a charge-breaking CP-conserving minimum.\\

In the first case, \eqref{chargeCons}, that is, $\cos(\theta) = \pm 1$, we have $\tvec{k}^\trans \tvec{k} =1$ at the vacuum. 
Also Froggatt and Nielsen require that the value of the potential at the minimum is independent of $\omega$, meaning that, with view on \eqref{Ks}, in the approach developed in the present work, this corresponds to a circle-shaped vacuum.
This in turn requires, as can be seen from table \ref{tabMPP}, that there is a basis where the parameters must satisfy
\begin{equation}
	M = E - u \unitmatrix_3 = \diag(0, 0, E_{33}-u) \quad
\text{yielding} \quad \lambda_5 = 0 \quad\text{and}\quad u = \frac{1}{4}\lambda_4\, .
\end{equation}

The value of $u$ being fixed, we may solve equations~\eqref{eqbord} and \eqref{cond12u}, giving
\begin{gather}
	\eta_3^2 = (\eta_{00} + u)(E_{33}-u)\,,
\end{gather}
which in terms of the conventional parameters gives
\begin{equation}
	\pm \sqrt{\lambda_1 \lambda_2} + \lambda_3 + \lambda_4 = 0\,,
\end{equation}
Requiring that $|k_3| < 1$, it can be shown that the choice of the solution depends on the sign of $\lambda_1$ and $\lambda_2$, thus giving
\begin{gather}
	\lambda_1 > 0\quad,\quad\lambda_2 > 0 \quad,\quad +\sqrt{\lambda_1 \lambda_2} + \lambda_3 + \lambda_4 = 0\quad,\quad \lambda_5 = 0 \label{minimum1}\\[.1cm]
	\textbf{or} \nonumber\\[.1cm]
	\lambda_1 < 0\quad,\quad\lambda_2 < 0 \quad,\quad -\sqrt{\lambda_1 \lambda_2} + \lambda_3 + \lambda_4 = 0\quad,\quad \lambda_5 = 0\,. 
	\label{minimum2}
\end{gather}
In addition, ensuring that the extremum is a minimum rules out the second solution \eqref{minimum2}. 
The only remaining set of constraints \eqref{minimum1} corresponds to the one derived in \cite{Froggatt:2004st}.\\

We now turn to the second case~\eqref{chargeBr} where $\cos(\theta) = 0$. Making use of~\eqref{Ks}, this means that $k_1 = k_2 = 0$ and, in the case 
where neither $\phi_1$ nor $\phi_2$ vanish, $|k_3| < 1$.\footnote{Note that for a vacuum at a large scale we cannot have $\phi_1 = \phi_2 = 0$.}
In any case, from a geometric point of view, the solution is a point, meaning that none of the eigenvalues of the matrix $E$ should vanish. 
Applying the constraints~\eqref{con} and~\eqref{cond12} gives:
\begin{equation}
	\eta_3^2 = E_{33} \eta_{00}
\end{equation}
or, in conventional parameters,
\begin{equation}
	\lambda_3 = \pm \sqrt{\lambda_1 \lambda_2}\,,
\end{equation}
where obviously the quantity $\lambda_1 \lambda_2$ must be positive. In addition to the above constraint, we can work out the condition $|\tvec{k}| < 1$ to give in the conventional formalism:
\begin{equation}
	\lambda_1 < 0\;,\; \lambda_2 < 0\;,\; \lambda_3 = \sqrt{\lambda_1 \lambda_2} \quad\textbf{or}\quad \lambda_1 > 0\;,\; \lambda_2 > 0\;,\; 
	\lambda_3 = -\sqrt{\lambda_1 \lambda_2} \, .
\end{equation}

Once again we want to ensure that the solution is a minimum, meaning here that $E_{33}$ must be positive. This rules out the first solution while the second set of conditions corresponds to the one in \cite{Froggatt:2004st}. Thus the present formalism agrees with the results of Froggatt and Nielsen in both cases.\\

Finally, relation~\eqref{cond2THDM} must hold if the MPP is to be realized. This results in an additional constraint among the beta-functions, 
that is\footnote{The fact that $\beta_{\eta_1}$ and $\beta_{\eta_2}$ vanish is due to the property of the Yukawa sector that each fermion couples to only one doublet.}:
\begin{equation}
	\beta_{\eta_{00}} + 2 k_3 \beta_{\eta_3} + k_1^2 \beta_{E_{11}} + k_2^2 \beta_{E_{22}} + k_3^2 \beta_{E_{33}} = 0\,.
\end{equation}
Injecting the expression of the bilinear couplings in terms of the conventional parameters gives:
\begin{equation}\label{betaFN}
	(1 + k_3)^2 \beta_{\lambda_1} + (1 - k_3)^2 \beta_{\lambda_2} + 2(1-k_3^2) \beta_{\lambda_3} + 2(k_1^2 + k_2^2)\beta_{\lambda_4} + 2(k_1^2 - k_2^2)\beta_{\lambda_5} = 0\,.
\end{equation}

In the case of the charge-conserving vacuum~\eqref{chargeCons} we have $\lambda_5 = \beta_{\lambda_5} = 0$ and $k_1^2 + k_2^2 + k_3^2 = 1$. 
Using these relations as well as the expression of $k_3$ in \eqref{cond12u} in terms of the conventional parameters we find
\begin{equation}
	\frac{1}{2}\sqrt{\frac{\lambda_2}{\lambda_1}} \beta_{\lambda_1} + \frac{1}{2} \sqrt{\frac{\lambda_1}{\lambda_2}} \beta_{\lambda_2} + \beta_{\lambda_3}  + \beta_{\lambda_4} = 0 \, .
\end{equation}
In the other case \eqref{chargeBr}, we have $k_1 = k_2 = 0$ and with \eqref{cond12} we find
\begin{equation}
	\frac{1}{2}\sqrt{\frac{\lambda_2}{\lambda_1}} \beta_{\lambda_1} + \frac{1}{2} \sqrt{\frac{\lambda_1}{\lambda_2}} \beta_{\lambda_2} + \beta_{\lambda_3} = 0 \, .
\end{equation}

These two expressions exactly match the condition (26) from ref.\ \cite{Froggatt:2004st}, with the only difference that here we did not need to consider the sign of $\lambda_4$.
 Instead, what distinguishes the two cases is the shape of the MPP vacuum.\\

To summarize this section, we have shown that the constraints derived by Froggatt and Nielsen \cite{Froggatt:2004st, Froggatt:2006zc} and 
later reused by McDowall and Miller \cite{McDowall:2018ulq} in the framework of a THDM type II constitute in fact a very particular case of the 
application of the formalism developed in this paper and can be easily recovered. 
We stress that numerous different realizations of the MPP may be derived in the same manner using the classification from table~\ref{tabMPP}, which 
might lead to various phenomenological implications of this principle at the EW scale. 
Let us emphasize that the conclusion in the works \cite{Froggatt:2004st, Froggatt:2006zc, McDowall:2018ulq}, that the MPP in the THDM cannot be realized 
for a SM-like Higgs-boson mass and the observed top-quark mass is based on a special case of the MPP.
Here we have seen in the geometric approach in terms of bilinears that the THDM may develop many more
different kinds of realizations of the MPP.

\subsection{Example potential}
\label{examplePotential}

As an additional example we study a CP conserving THDM potential 
in which the parameters in conventional notation \eqref{Vconv}
satisfy
\begin{equation}
\lambda_1 = \lambda_2 = \lambda_3, \quad 
\lambda_4 = \lambda_5 = \lambda_5^*, \quad
\lambda_6 = \lambda_6^* = \lambda_7, \quad 
\text{ with } \lambda_4 \neq - \lambda_5.
\end{equation}
Let us recall that these relations are assumed to hold at the scale $\Lambda$.
In bilinear space \eqref{pot} this corresponds
to the quartic parameters in the form
\begin{equation}
E = \diag (E_{11}, 0, 0 ), \text{ with } E_{11} \neq 0, 
\qquad
\tvec{\eta} = \begin{pmatrix} \eta_1, & 0, & 0 \end{pmatrix}^\trans, \text{ with } \eta_1 \neq 0.
\end{equation}
With view on table \ref{tabMPP} we have the case $\det(E)=0$, 
$\tr^2(E)-\tr(E^2)=0$, but with $\tr(E) \neq 0$.
together with $\tvec{\eta}^\trans E \tvec{\eta} = E_{11} \eta_1^2 \neq 0$
and also 
$(E \tvec{\eta} \times \tvec{\eta})^\trans (E \tvec{\eta} \times \tvec{\eta}) = 0$,
that is the MPP is realizable as a disk supposed $\eta_{00}$ satisfies 
the condition \eqref{cond12}. 

Let us look into the solutions in detail. 
First we note that we can, by a change of basis \eqref{b2}, with the rotation matrix 
\begin{equation}
R = 
\begin{pmatrix}
0 & \phantom{+}0 & - 1\\
0 & \phantom{+}1 & \phantom{+}0\\
1 & \phantom{+}0 & \phantom{+}0
\end{pmatrix}
\end{equation}
shift both,
the diagonal entry as well as the corresponding entry of $\tvec{\eta}$.
This case corresponds therefore to the case \eqref{2zero} with 
two vanishing eigenvalues of $E$.

In order to study the MPP we consider first the case $|\tvec{k}|<1$. 
The second vacuum follows from \eqref{con} 
\begin{equation} \label{vac21}
\langle  \tvec{k} \rangle_2 = 
\begin{pmatrix} -\frac{\eta_1}{E_{11}}, & y, & z \end{pmatrix}^\trans
\quad \text{with } y^2 + z^2  < 1 -  \frac{\eta_1^2}{E_{11}^2} 
\text{ for } \eta_1^2 < E_{11}^2.
\end{equation}
This is indeed an open disk in the $y-z$ direction with 
border radius $\sqrt{  1 - \eta_1^2/E_{11}^2}$.
The parameter $\eta_{00}$ has thereby to fulfill the 
condition \eqref{cond12},
\begin{equation} \label{sol1eta00}
\eta_{00} =  \frac{\eta_1^2}{E_{11}}\;.
\end{equation}
Moreover, the $\beta$ functions have to satisfy \eqref{cond2THDM}.
In the case that we only consider the potential without any Yukawa couplings 
these conditions read
\begin{equation}  \label{exabeta}
\beta_{\eta_{00}} =  2 \frac{\eta_1}{E_{11}} \beta_{\eta_1}  - \left( \frac{\eta_1}{E_{11}}\right)^2 \beta_{E_{11}}
\end{equation}
again at the scale $\Lambda^2$. 

For $|\tvec{k}|=1$ the stationarity condition is 
given by \eqref{eqbord} with a Lagrange multiplier $u$. 
For $u=0$ we get an exceptional solution at the border of
the solution \eqref{vac21},
\begin{equation} \label{vac21u}
\langle  \tvec{k} \rangle_2 = 
\begin{pmatrix} -\frac{\eta_1}{E_{11}}, & y, & z \end{pmatrix}^\trans
\quad \text{with } y^2 + z^2  = 1 -  \frac{\eta_1^2}{E_{11}^2}
\text{ for } \eta_1^2 \le E_{11}^2.
\end{equation}
which is, as to be expected from table \ref{tabMPP} a circle
in case that $\eta_{00}$ fulfills \eqref{cond12u} which equals \eqref{sol1eta00}
in this case. Also the $\beta$ functions have to fulfill \eqref{cond2THDM}
which give, neglecting Yukawa interactions \eqref{exabeta}.

For $u \neq 0$ we immediately get the solution from \eqref{eqbord}
\begin{equation} \label{vac22}
 \langle  \tvec{k} \rangle_2 = 
\begin{pmatrix}  \pm 1  & 0, & 0 \end{pmatrix}^\trans ,
\end{equation}
that is, two possible points 
with corresponding values for the Lagrange multiplier 
\begin{equation}
u = E_{11} \pm \eta_1 \;.
\end{equation}
The condition \eqref{cond12u} restricts the parameter $\eta_{00}$, 
\begin{equation} \label{soleta002}
\eta_{00} = - E_{11} \mp 2 \eta_1 \;.
\end{equation}
corresponding to one of the two discrete vacua in \eqref{vac22}.
Besides, we have to satisfy the condition for the beta functions \eqref{cond2THDM},
that is,
\begin{equation} \label{ex1beta}
\beta_{\eta_{00}}  \pm  2 \beta_{\eta_{1}}   + 
\beta_{E_{11}}  = 0.
\end{equation}

We note that the potential is CP conserving \cite{Maniatis:2007vn} (see section \ref{sec:bilinears})
and we conclude that the MPP is realizable 
in the stronger sense with a continuous disk of degenerate stationary points with 
the parameters and its $\beta$ functions satisfying the discussed constraints. 
If these constraints are not fulfilled we may get at most an isolated point, that is,
the MPP in the weaker sense, where the parameters and $\beta$ functions
 have to have in particular
to satisfy \eqref{soleta002} and \eqref{cond2THDM}.

\section{Application of the MPP: From the high scale to the EW scale}
\label{analysisMPP}

Having classified the possible types of vacua at the high scale $\Lambda$, we now want to study the MPP and its low-energy phenomenological implications. 
The method we use in this analysis can be summarized by the following steps:
\begin{itemize}
	\item At the high scale $\Lambda$ we encounter 7 real parameters from the quartic part of the potential besides
	the parameters of the Yukawa couplings. The three gauge couplings can be run up from their known values
	at the electroweak scale. 
	
	\item We consider the constraints provided by the MPP at the high scale $\Lambda$, as given by Tab.~\ref{tabMPP}.
	
	\item We run all couplings down to the electroweak scale by the evolution equations at one-loop accuracy. 
	At the electroweak scale we have to consider in addition the 
	quadratic parameters of the potential. These quadratic parameters are constrained since the model should provide
	 the observed spontaneous electroweak symmetry breaking.
	
	\item 
	We scan over all remaining free parameters.
	
	\item For every parameter set, we compute the masses of the physical Higgs bosons of the THDM and the masses of the fermions.
\end{itemize}

The purpose of this analysis is to study the implications of the MPP on the masses of the physical states. 
The main goal is to determine whether the application of the MPP to the THDM may yield correct (\textit{i.e.} observed) masses of a Standard-Model-like Higgs boson and the top-quark. 
In appendix \ref{EWbreak} we recall the mechanism of spontaneous symmetry breaking in the THDM in
the bilinear formalism. In particular we present in this appendix the mass matrices of the Higgs bosons and
the mass of the pair of charged Higgs bosons.

\subsection{First example: MPP as a spherical vacuum}

As an application of our methods, 
we first study a THDM potential with the MPP realized as 
a spherical vacuum characterized by the potential parameter matrix $E=0$ (respectively $M=0$ for $|\tvec{k}|=1$) (see table~\ref{tabMPP}). We consider for simplicity only the top Yukawa coupling and we stay in the framework of a THDM type III to keep the discussion general. Note however that other types of THDM may always be obtained as special cases when either $y_t$ or $\epsilon_t$ is set to 0 in some well-chosen basis. In appendix~\ref{top} we show the Lagrangian of the top quark Yukawa coupling and its behavior under Higgs-basis changes.

Following the discussion in section~\ref{class} we have to consider the cases $|\tvec{k}| < 1$
and $|\tvec{k}| = 1$. In the first case, equation~\eqref{nullDK0} reads
\begin{align}
\begin{split}
	g \tvec{k}^\trans \tvec{k} + G &= \frac{3}{2}\left[ Y^\dagger\left(\sigma_0 + \tvec{k}^\trans \tvec{\sigma}\right) Y \right]^2\\
	&= \frac{3}{2}\left[(1+k_3) \abs{y_t}^2 + (1-k_3)\abs{\epsilon_t}^2 + k_1(\epsilon_t^* y_t + y_t^*\epsilon_t) + i k_2(\epsilon_t^* y_t - y_t^*\epsilon_t)\right]^2\,. \label{sphereCase1}
\end{split}
\end{align}

In order to simplify the evaluation of this relation, we may first perform a change of basis making the top quark Yukawa couplings real. Defining $y_t = \abs{y_t} e^{i \theta_y}$ and $\epsilon_t = \abs{\epsilon_t} e^{i \theta_\epsilon}$, the associated unitary transformation is (see appendix \ref{top})
\begin{equation}
\label{Ureal}
	U = \begin{pmatrix}
		e^{-i \theta_y} & 0 \\
		0 & e^{-i \theta_\epsilon}
	\end{pmatrix}\,,
\end{equation}
and corresponds, in terms of bilinears, to a rotation matrix $R(U)$ \eqref{unitR}. The latter is in fact a rotation around the $z$-axis which in our case can be performed without loss of generality. Equation \eqref{sphereCase1} can be further simplified after rotation to a basis where \mbox{$Y = (y'_t, \epsilon'_t)^\trans = (y'_t,  0)^\trans$}. The associated transformation $U$ is a 2D rotation matrix and corresponds, in the bilinear space, to a rotation around the $y$-axis. In this new basis, \eqref{sphereCase1} now reads 
\begin{equation}
	g \tvec{k}^\trans \tvec{k} + G = \frac{3}{2} (1+k_3)^2 y_t^4\,,
\end{equation}
where we dropped the primes for clarity. We see that choosing a point $\tvec{k}$ within the sphere fixes the value of the top Yukawa coupling. Furthermore, it can be shown that a necessary condition for the above equation to be satisfied is\footnote{To obtain this inequality, we used the fact that $g < G$ at all energy scales.} 
\begin{equation}
	\abs{y_t} \geq \left(\frac{g + G}{6} \right)^{1/4}\, ,
\end{equation}
which can be reformulated in a basis-invariant way:
\begin{equation}
	\label{yukInequality}
	\sqrt{\abs{y_t}^2 + \abs{\epsilon_t}^2} \geq \left(\frac{g + G}{6}  \right)^{1/4} \approx 0.38 \quad \text{for} \quad \Lambda=\SI[parse-numbers=false]{10^{18}}{\GeV}\, .
\end{equation}

Given the above inequality, and under the assumption that the left hand-side is roughly of order $\abs{y_t^{SM}}$, we expect the top quark to be the only fermion that couples strongly enough to the Higgs doublets to satisfy it. However the situation may not be as clear in some limiting cases, e.g. the THDM type II with a high value of $\tan \beta$.\\

We now turn to the case $|\tvec{k}| = 1$, where the evaluation of~\eqref{nullDK0u} in the same basis as above gives a constraint involving the Lagrange multiplier $u$, namely 
\begin{equation}
	16 u^2 + g + G = \frac{3}{2} (1+k_3)^2 y_t^4\, .
\end{equation}
We note that in this case, condition \eqref{yukInequality} must be satisfied as well.

Choosing a specific value for $u$ and $k_3$ we can fix all the relevant parameters at the high scale: In case $|\tvec{k}| < 1$ all the quartic couplings vanish whereas in case $|\tvec{k}| = 1$ the non-zero parameters are related to the Lagrange multiplier $u$ via
\begin{subequations}
	\begin{align}
		\eta_{00} &= -u\,,\\
		E_{ii} &= u\:\:,\:\: i = 1,2,3\,.
 	\end{align}
\end{subequations}

It is remarkable that in both cases this set of couplings implies CP conservation at the level of the quartic part of the scalar potential \cite{Maniatis:2006fs}. The only remaining possible source of CP violation is a non-zero value for the scalar mass coupling $\xi_2$ in the basis chosen above.\\

The next step is to perform the running of the couplings down to the electroweak scale, where the study of spontaneous symmetry breaking will eventually allow for the determination of the masses of the Higgs bosons and the fermions. 
At the electroweak scale we have to consider the quadratic parameters $\xi_0$, $\xi_{i,\ i=1,2,3}$  of the Higgs potential.
These couplings are however subject to constraints in order to give a proper $SU(2)_L\times U(1)_Y\rightarrow U(1)_{em}$ symmetry breaking pattern.
 Using~\eqref{EWbasisChange}, we can trade these four parameters for $u_\text{EW}$, $\beta$, $\zeta$ and $v_0$. The latter is known since it corresponds to the Standard Model vacuum-expectation value $v_0 \approx \SI{246}{\GeV}$.\\

The angles $\beta$ and $\zeta$ 
parametrize the basis transformation \eqref{EWbasisChange} which allows to achieve the form of the Higgs basis \eqref{higgsBasis}.
We note that a non-zero value for $\zeta$ will in our case generate CP violation at the level of the scalar potential, and accordingly imply a mixing between the scalar and pseudo-scalar physical states. For simplicity reasons
we consider the case $\zeta = 0$, in which we will identify the lightest neutral scalar
with the CP-even Standard-Model like Higgs boson.\\

\subsection{Second example: MPP as a disk-shaped vacuum}

We now briefly discuss the next-to-maximal symmetric vacuum, namely the disk-shaped one. This includes in particular the CP-conserving potential discussed in section~\ref{examplePotential}.

In the following, we will work in a basis where $E = \diag (0, 0, E_{33})$, respectively $M = \diag (0, 0, M_{33})$ if $\abs{\tvec{k}} = 1$. In the latter case, the MPP vacuum reduces to a circle. Using the results from section~\ref{statPoints}, the quartic parameters at the high scale satisfy the following constraints in order to provide stationary points:
\begin{align}
	\abs{\tvec{k}} < 1 &: \begin{cases}
		\eta_3 + E_{33} k_3 = 0 \, ,\\
		\eta_{00} + \eta_3 k_3 = 0\, ,\\
		k_1^2 + k_2^2 < 1 - k_3^2\, ,
	\end{cases}\\[.15cm]
	\abs{\tvec{k}} = 1 &: \begin{cases}
		\eta_3 + M_{33} k_3 = 0\, , \\
		u + \eta_{00} + \eta_3 k_3 = 0\, ,\\
		k_1^2 + k_2^2 = 1 - k_3^2\, .
	\end{cases}
\end{align}
Note that, as discussed earlier, all other quartic parameters have to vanish at the high scale. The constraints from the quantum potential \eqref{nullDK0}, \eqref{nullDK0u} respectively simplify as
\begin{gather}
\begin{split}
\label{diskQpot1}
	8 E_{33}^2 \left(1 - k_3^2 \right)^2 + g\tvec{k}^\trans \tvec{k} + G &= \frac{3}{2}\left[ Y^\dagger\left(\sigma_0 + \tvec{k}^\trans \tvec{\sigma}\right) Y \right]^2 \, ,
\end{split}\\
\begin{split}
\label{diskQpot2}
	16 u^2 + 8 M_{33}^2 \left(1 - k_3^2 \right)^2 + g + G &= \frac{3}{2}\left[ Y^\dagger\left(\sigma_0 + \tvec{k}^\trans \tvec{\sigma}\right) Y \right]^2 \, .
\end{split}
\end{gather}

Although it is still possible to make the Yukawa couplings real using transformation~\eqref{Ureal}, we cannot rotate to a basis where either $y_t$ or $\epsilon_t$ vanishes without introducing a mixing between $k_1$ and $k_3$. In this context, equations~\eqref{diskQpot1} and~\eqref{diskQpot2} can be respectively rewritten as
\begin{gather}
\begin{split}
	8 E_{33}^2 \left(1 - k_3^2 \right)^2 + g\tvec{k}^\trans \tvec{k} + G &= \frac{3}{2}\left[(1+k_3) y_t^2 + (1-k_3)\epsilon_t^2 + 2 k_1 \epsilon_t y_t\right]^2\, ,
\end{split}\\
\begin{split}
	16 u^2 + 8 M_{33}^2 \left(1 - k_3^2 \right)^2 + g + G &= \frac{3}{2}\left[(1+k_3) y_t^2 + (1-k_3)\epsilon_t^2 + 2 k_1 \epsilon_t y_t\right]^2\, .
\end{split}
\end{gather}

Note that the number of free parameters at the high scale is increased here compared to the spherical case. We also emphasize that the previous comments about CP conservation are still valid in this case.

\subsection{Results and discussion}

We now present the results of the numerical analyses for the MPP for 
a Higgs potential providing a spherical and disk-shaped vacuum. 
In order to detect the minima at the electroweak scale we have to solve the 
corresponding gradient equation (see appendix~\ref{EWbreak} and \cite{Maniatis:2006fs}). In principle, we may encounter regular and
irregular solutions of these equations. In the spherical case, the irregular solutions provide very low values of $u_{EW}$. Consequently, either the scalar mass matrix \eqref{scalarMasses} develops negative eigenvalues or leads to situations where the lightest scalar is much lighter than the SM-like Higgs boson. On the other hand, in the disk and circle cases, these solutions may provide some acceptable values for the masses of the physical states. In any case, for simplicity, we consider here only the regular solutions of eq.~\eqref{eqEW}.

\begin{figure}[h]
	\centering
	\begin{subfigure}[h]{0.49\linewidth}
         \centering
         \includegraphics[width=\textwidth]{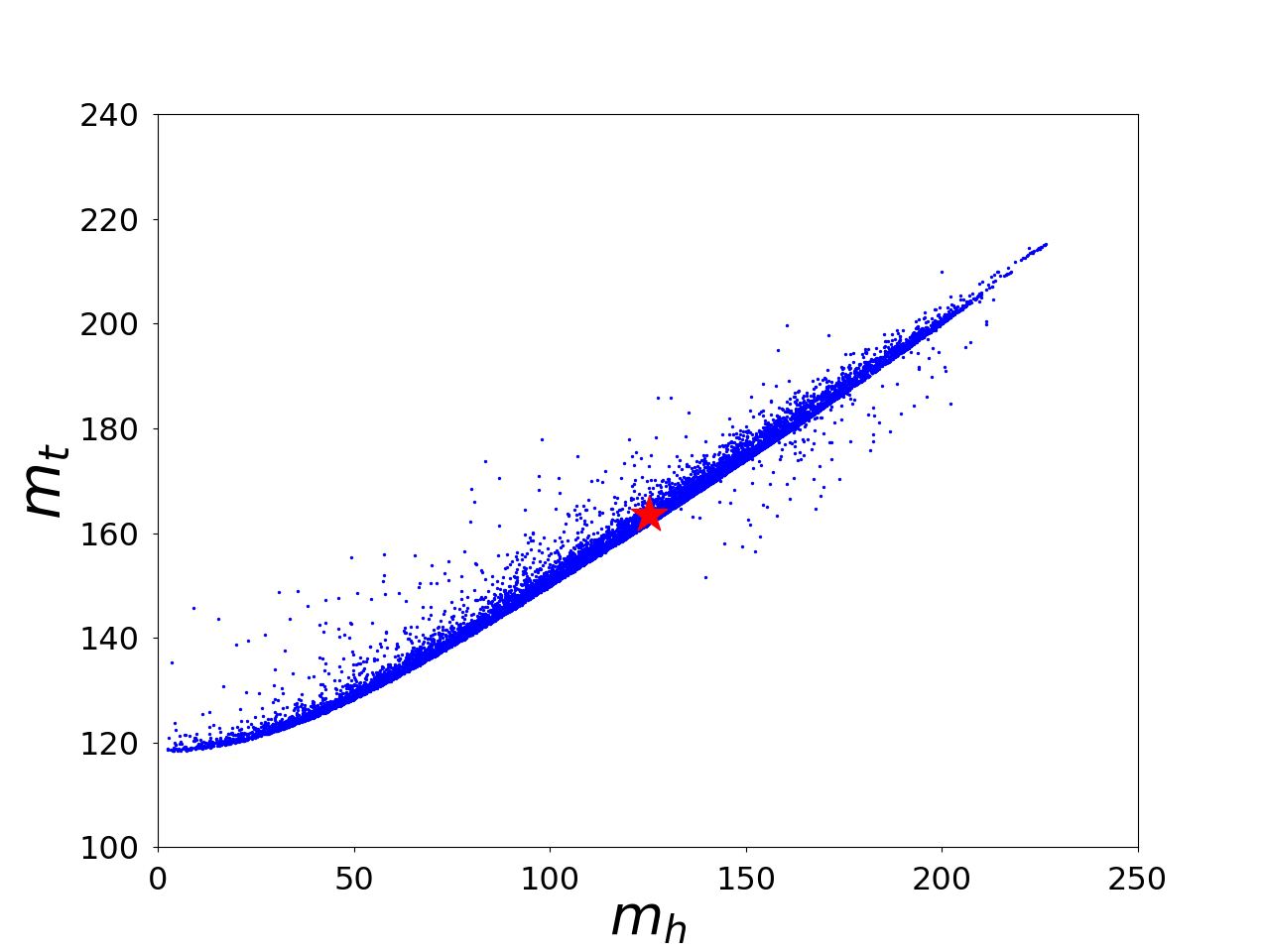}
	\end{subfigure}
	\hfill
	\begin{subfigure}[h]{0.49\linewidth}
         \centering
         \includegraphics[width=\textwidth]{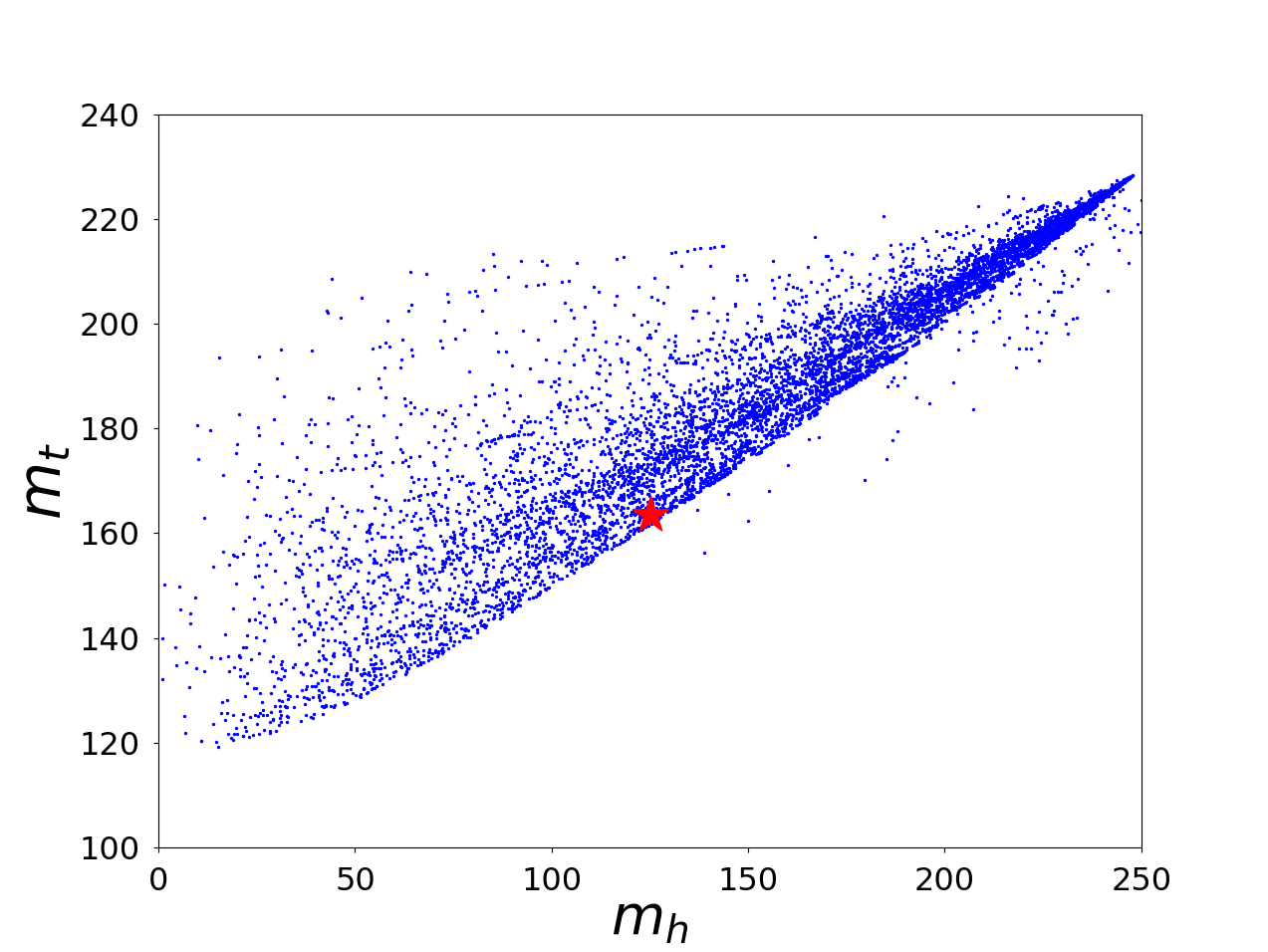}
	\end{subfigure}\\
	\begin{subfigure}[h]{0.49\linewidth}
         \centering
         \includegraphics[width=\textwidth]{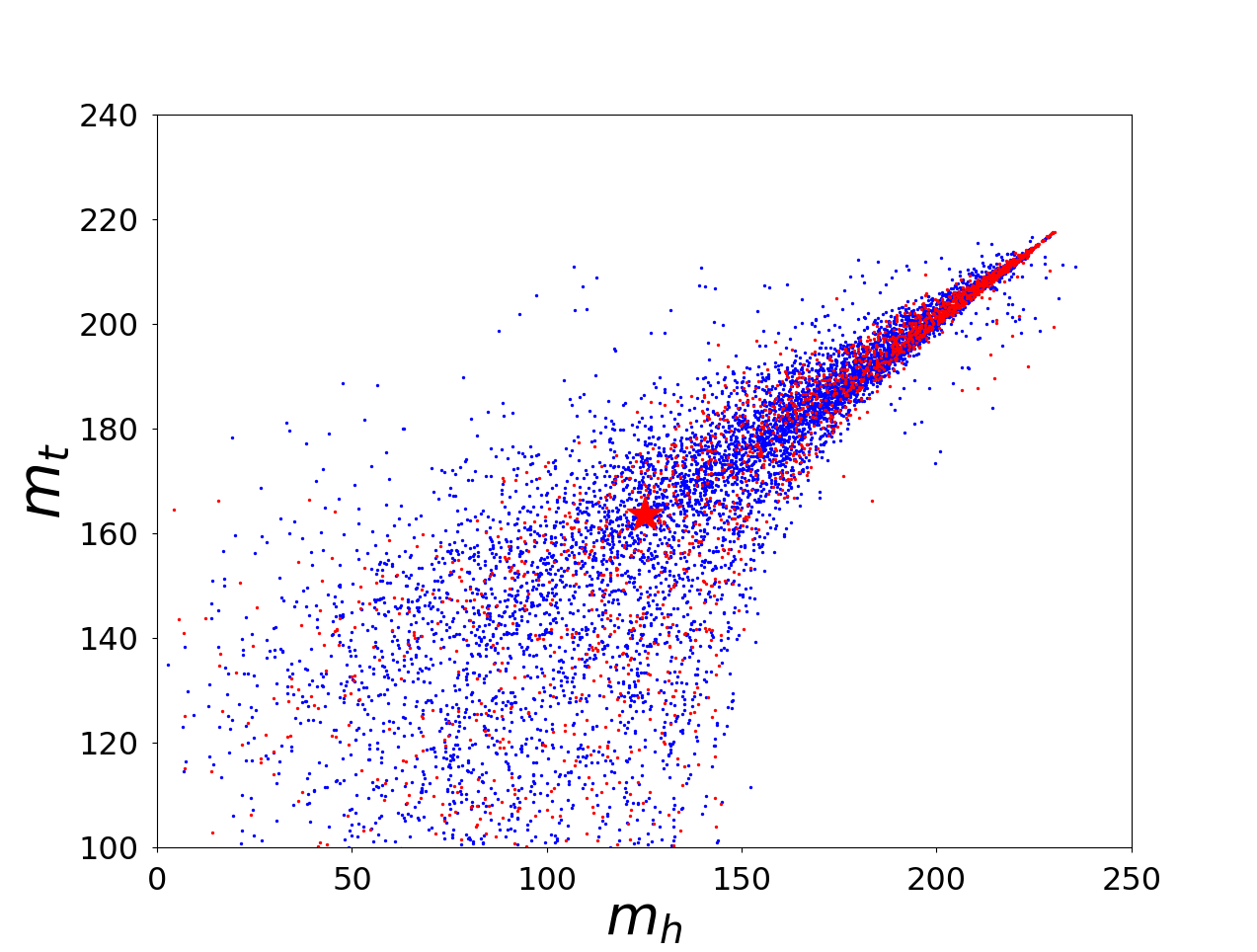}
         \caption{}
	\end{subfigure}
	\hfill
	\begin{subfigure}[h]{0.49\linewidth}
         \centering
         \includegraphics[width=\textwidth]{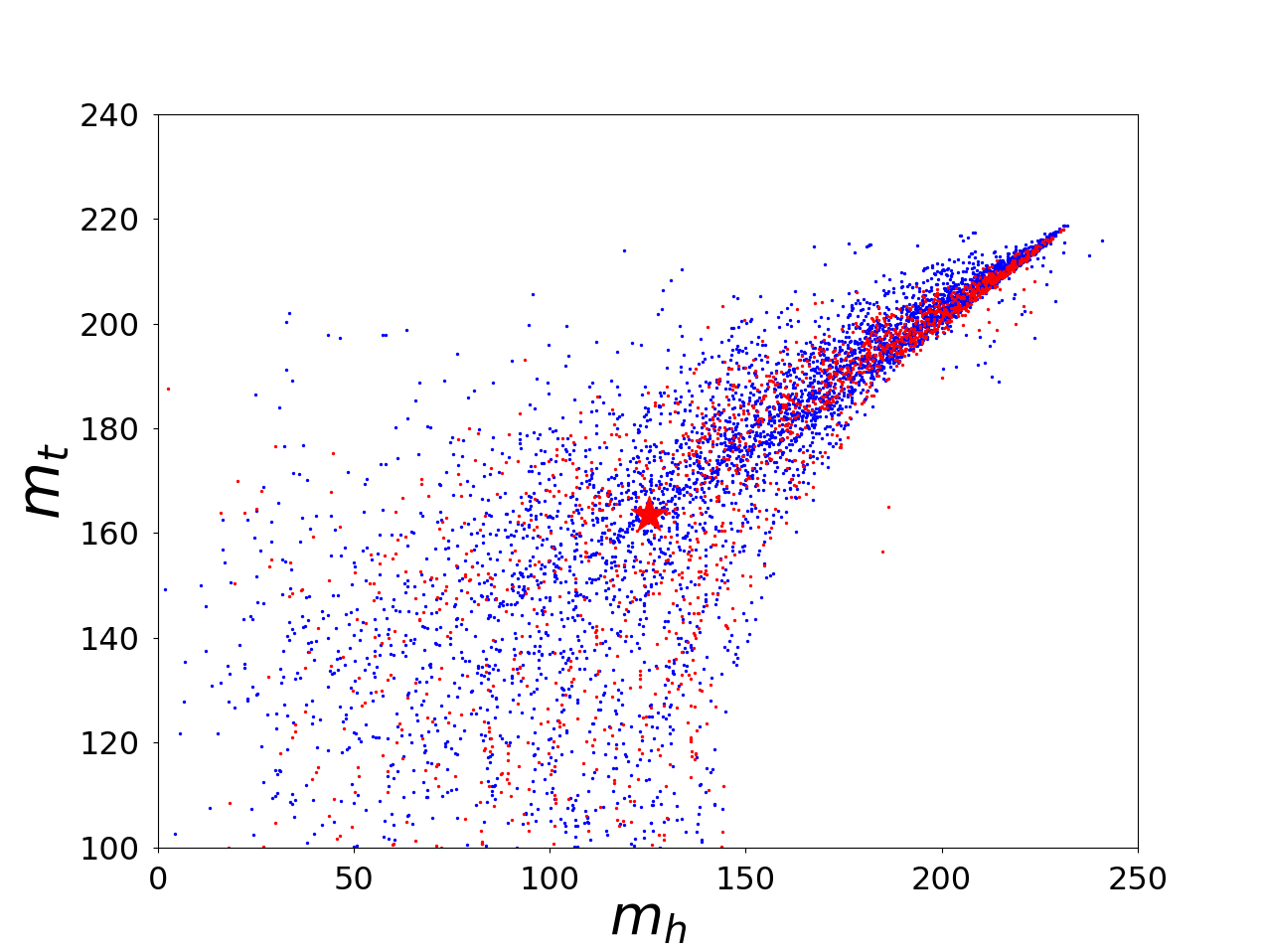}
         \caption{}
	\end{subfigure}
	\caption{Results of the parameter scans in the $(m_h, m_t)$ plane in cases (a) $|\tvec{k}| < 1$ and (b) $|\tvec{k}| = 1$ for $\Lambda = \SI[parse-numbers=false]{10^{18}}{\GeV}$. The two upper plots are obtained in the case of a spherical vacuum, and the lower ones in the case of a disk/circle-shaped vacuum. For the points shown in red, the scalar potential is bounded from below at all energy scales between $M_Z$ and $\Lambda$. The red stars indicate the position of $(m_h, m_t^{\overline{\mathrm{MS}}}) \approx (\SI{125}{\GeV}, \SI{163}{\GeV})$. \label{BManalysis}}
\end{figure}

For these solutions, we show in  Fig.~\ref{BManalysis} the results of the numerical analysis in the $(m_h, m_t)$ plane for $\Lambda = \SI[parse-numbers=false]{10^{18}}{\GeV}$. The list of free parameters at the high scale along with their allowed range is presented in Table~\ref{freeParams}. We systematically excluded from the parameter scan the points violating perturbativity and unitarity bounds \cite{Branco:2011iw}. We also studied the conditions of a bounded from below potential \cite{Branco:2011iw, McDowall:2018ulq} at all energy scales as illustrated in Fig.~\ref{BManalysis}. We finally note that the ranges of $u$, $y_t$ and $\epsilon_t$ at the high scale were chosen based on the observations that $\abs{u} > 0.5$ sytematically violate perturbativity and unitarity bounds and that for Yukawa couplings greater than $1$ the RG flow tends to develop Landau poles.

At the EW scale, we have to scan over two remaining free parameters, namely $\beta$ and $u_{EW}$. The former is taken in the range $[\shortminus\frac{\pi}{2}, \frac{\pi}{2}]$ and the latter\footnote{In order to give a physical meaning to $u_{EW}$, we note that $m_{H^\pm} =v_0\sqrt{ 2 u_{EW} }$ takes values between $0$ and $\SI{1.1}{\TeV}$ when $0 < u_{EW} < 10$.} in the range $[0, 10]$. We emphasize that the classical potential is ensured to develop a global minimum at the EW scale since we require Theorem 3 of \cite{Maniatis:2006fs} to be satisfied.

\begin{table}[h]
{
\hfill
\begin{tabular}{|C{1cm}|C{1.5cm}|C{1.5cm}|}
\hline
\multicolumn{3}{|c|}{\large Spherical vacuum}                                       \\ \hline
 $\abs{\tvec{k}}$ & $\left[0, 1\right[$ & $1$                                 \\ \hline
 $u$              & /                   & $\left[\shortminus 0.5, 0.5\right]$ \\ \hline
 $k_3$            & $\left]\shortminus1, 1\right[$ & $\left[\shortminus1, 1\right]$                 \\ \hline
\end{tabular}
\hfill
\begin{tabular}{|C{1cm}|C{1.5cm}|C{1.5cm}|}
\hline
\multicolumn{3}{|c|}{Disk-shaped vacuum}                                                \\ \hline
$\abs{\tvec{k}}$ & $\left[0, 1\right[$            & $1$                                 \\ \hline
$u$              & /                              & $\left[\shortminus 0.5, 0.5\right]$ \\ \hline
$k_3$            & $\left]\shortminus1, 1\right[$ & $\left[\shortminus1, 1\right]$      \\ \hline
$k_1$            & $\left]\shortminus1, 1\right[$ & $\left[\shortminus1, 1\right]$      \\ \hline
$y_t$            & \multicolumn{2}{c|}{$\left[0, 1\right]$}                             \\ \hline
$\epsilon_t$     & \multicolumn{2}{c|}{$\left[0, 1\right]$}                             \\ \hline
\end{tabular}
\hfill
}
\caption{Ranges of the free parameters at the high scale, in the spherical and disk case. The value of the $k_i$'s are always chosen such that $\abs{\tvec{k}} \leq 1$.}
\label{freeParams}
\end{table}

In all four considered cases, there are regions of the parameter space for which the MPP constraints are compatible with the observed values of $m_h$ and $m_t$. However, if we also take the constraint into account that the potential has to be bounded from below, the spherical vacua are ruled out. In contrast, the disk- and circle-shaped vacua provide acceptable values for the masses along with the requirement of a stable potential. Although the question of identifying the precise regions of the parameter space yielding correct $(m_h, m_t)$ values goes beyond the scope of this work, the conclusion of this simple analysis is that the realization of the MPP is \textit{a priori} not incompatible with a SM-like Higgs boson and top quark in the context of the general THDM.

Eventually, we would like to mention that we varied in our numerical analysis the value of the high scale $\Lambda$ in the range $10^{15} - \SI[parse-numbers=false]{10^{20}}{\GeV}$ and found that our predictions on the masses show a negligible dependence on this parameter.

\section{Conclusion and outlook}

The MPP \cite{Froggatt:1995rt} forces the Higgs potential to provide degenerate vacua with the same potential value.
A long time before the discovery of the Higgs boson, its mass has been predicted to be $135 \pm 9 \text{ GeV}$
based on this principle applied to the Standard Model -- a quite remarkable result.\\
 
Some effort \cite{Froggatt:2004st, Froggatt:2006zc, McDowall:2018ulq}
has been spent to apply the MPP to the TDHM. 
This has been done in the conventional 
formalism, where the gauge degrees of freedom appear explicitly. Here we have studied the MPP
in the THDM applying the bilinear formalism \cite{Nagel:2004sw, Maniatis:2006fs, Nishi:2006tg} . 
This formalism turns out to be quite powerful to 
study models with additional Higgs-boson doublets.\\

We have classified all different types of degenerate vacua 
 in the THDM. In particular, we find degenerate vacua which realize the MPP in a weaker sense, providing
additional isolated points, but also realizations in  a stronger sense of line segments, circles, 
surfaces of spheres, as well as spheres.
We have presented the classification in a basis-invariant way.
For any THDM the corresponding conditions for the different types of realizable MPP's can be easily checked. \\

We have studied the $\beta$-functions of the THDM in detail in terms of bilinears and have shown that the MPP, considering only the THDM potential, is not realizable.
This changes if we consider the Yukawa couplings.\\

We recover the MPP cases studied in the literature but in addition can identify
different realizations of the MPP in the THDM. 
We explored in section~\ref{analysisMPP} four different realizations of the MPP in the context of a general THDM with a simplified matter content. Two of them (the disk- and circle-shaped vacua) are compatible with the measured values of the Higgs and top quark masses, satisfying simultaneously the constraints of a correct electroweak symmetry breaking, perturbativity and unitarity as well as the requirement of a stable scalar potential at all energy scales.\\




Our analysis is done at the one-loop order of the $\beta$-functions and using the tree-level RG-improved potential. In the future we would like to extend this analysis to higher orders of the effective potential, the RGEs, as well as the matching conditions between the $\overline{\mathrm{MS}}$ and pole masses. Also we would like to consider the experimental constraints from negative searches of additional Higgs bosons. Eventually, we plan to extend this study by considering all three families of fermions also in the context of THDMs with natural flavour conservation.


\acknowledgments
The project was supported in part by the UBB projects ''Materia Obscura y los bosones de Higgs''
with No. DIUBB 193209 1/R and Fondecyt with No. 1200641.

\appendix

\section{Electroweak symmetry breaking in the THDM}
\label{EWbreak}

Here we briefly give some essential details of electroweak symmetry breaking in the THDM 
in terms of bilinears \cite{Maniatis:2006fs}. 

At the minimum of the potential, the Higgs doublet expectation values can be expressed as:
\begin{align} \label{min}
    \langle \varphi_1 \rangle = \begin{pmatrix} v_1^+ \\ v_1^0 \end{pmatrix} \ \ , \ \ \langle \varphi_2 \rangle = \begin{pmatrix} v_2^+ \\ v_2^0 \end{pmatrix}
\end{align}
with all the vacuum-expectation values being in general complex. Assuming a charge-conserving vacuum, a $SU(2)_L\times U(1)_Y$ transformation allows one to rewrite the fields in the form:
\begin{equation} \label{EWphis}
    \langle \varphi_1 \rangle  = \begin{pmatrix} 0 \\ |v_1^0| \end{pmatrix} \ \ , \ \ \langle \varphi_2 \rangle  = \begin{pmatrix} 0 \\ |v_2^0|\, e^{i \zeta} \end{pmatrix} 
    \, .
\end{equation}
After a basis transformation \eqref{EWbasisChange}, the vacuum-expectation values of the doublets at the minimum of the potential can be expressed as\footnote{This basis is generally referred to as the Higgs basis.} 
\begin{align} 
    \langle \varphi_1 \rangle = \frac{1}{\sqrt{2}}\begin{pmatrix} 0 \\ v_0 \end{pmatrix} \ \ , \ \ \langle \varphi_2 \rangle  = \begin{pmatrix} 0 \\ 0 \end{pmatrix} \ ,
\end{align}
where $v_0 \approx 246\, \mathrm{GeV}$ is the Standard Model vacuum-expectation value. 
Then we can expand the fields about the minimum, giving
\begin{align}\label{higgsBasis}
    \varphi_1(x) = \frac{1}{\sqrt{2}}\begin{pmatrix} 0 \\ v_0 + \rho'(x) \end{pmatrix} \ \ , \ \ \varphi_2(x)  = \begin{pmatrix} H^+(x) \\ \frac{1}{\sqrt{2}}\big(h'(x) + i h''(x) \big)  \end{pmatrix} \ .
\end{align}
As has been shown in \cite{Maniatis:2006fs}, the observed electroweak symmetry breaking,
that is, a non-trivial vacuum with both charged components of the doublets vanishing, \eqref{min},
corresponds to 
\begin{equation} \label{EW}
K_0^2 = \tvec{K}^\trans \tvec{K} \;.
\end{equation}
This minimum of the potential can be found from the gradient of the potential 
by introducing a Lagrange multiplier $u_\text{EW}$ in order to satisfy \eqref{EW}, that is,
from
\begin{equation} \label{eqEW}
\nabla \left( V  - u_\text{EW} (K_0^2 - \tvec{K}^\trans \tvec{K} ) \right) =0
\end{equation}
with $V$ the potential as given in \eqref{pot}.
Supposed the potential has the correct electroweak symmetry breaking, corresponding to 
a solution of \eqref{eqEW} the mass matrix
 for the neutral scalars $(\rho', h', h'')$ is given by:
\begin{align}
\begin{split}\label{scalarMasses}
    \mathcal{M}_{neutral} ={}& 2\begin{pmatrix} -\xi_0' - \xi_3' && -\xi_1' && -\xi_2' \\ -\xi_1' && v_0^2 (u_\text{EW} + E'_{11}) && v_0^2\, E'_{12} \\ -\xi'_2 && v_0^2\, E'_{12} && v_0^2 ( u_\text{EW} + E'_{22})\end{pmatrix} \\
    ={}& 2 v_0^2 \begin{pmatrix} \eta'_{00} + 2\eta'_3 + E'_{33} && \eta'_1 + E'_{13} && \eta'_2 + E'_{23} \\ \eta'_1 + E'_{13} && u_\text{EW} + E'_{11} &&  E'_{12} \\ \eta'_2 + E'_{23} &&  E'_{12} && u_\text{EW} + E'_{22}\end{pmatrix}\ ,
\end{split}
\end{align}
where the second equality was obtained using relation \eqref{eqEW}. Due to the dependence on $u_{\text{EW}}$, it appears that one way to approach the decoupling limit is to have high values for $u_\text{EW}$.

The mass of the charged Higgs is given by:
\begin{align}
    m_{H^\pm}^2 = 2 u_\text{EW}\, v_0^2 \ ,
\end{align}
where we see that the charged Higgs-boson mass squared is proportional to 
the Lagrange multiplier $u_\text{EW}$.


\section{Suppression of the quadratic terms of the THDM potential}
\label{appendixJ4}

Here we briefly argue that the MPP applied to the THDM  \eqref{eq-vk}
\begin{equation}
\label{avk}
V^{\text{THDM}} = K_0\, J_2(\tvec{k})+ K_0^2\, J_4(\tvec{k})
\end{equation}
requires that the function $J_4$ vanishes at a high scale $\Lambda$. 

We first note that $J_2(\tvec{k})$ and $J_4(\tvec{k})$
depend linearly on the quadratic and quartic potential parameters; see \eqref{J2J4},
 in addition to the dimensionless fields $\tvec{k}$
with $|\tvec{k}| \le 1$. Therefore we expect that the absolute
values of $J_2$ and $J_4$ are not much larger than one
since the parameters should not be too large for perturbativity 
reasons. 

We first consider the potential at the electroweak scale, that 
is at the scale $\Lambda={\cal O} (100\text{ GeV})$. The non-trivial
minimum of the potential is at $\langle K_0 \rangle_1= - J_2/(2 J_4)$ with
a corresponding potential value of $V^{\text{THDM}}  = - J_2^2/(4 J_4)$.
The MPP (see \eqref{Cond1} and \eqref{Cond2})
requires to have the same potential value at the high scale. 
We expect from the running of the parameters that also the functions
$J_2$ and $J_4$ depend on the scale, so let us denote
these functions at the high scale with a prime symbol, $J_2'$ and $J_4'$.
Even that they are in general different from $J_2$ and $J_4$ at the electroweak
scale, we expect their absolute values also not to be much larger than one. 

Now the condition, that the potential value is degenerate at the high scale gives
\begin{equation}
J_4' = \frac{- J_2^2 - 4 J_2' J_4 \Lambda^2}{4 J_4 \Lambda^4} \;.
\end{equation}
This in turn means that $J_4'$ goes to zero for large $\Lambda$
supposed $J_2$, $J_4$, and $J_2'$ are not too large.
This condition simply arises from the principle to have degenerate
vacua with the same potential value.

\section{One-loop RGEs in the bilinear formalism}
\label{rges}

The RGEs of the THDM were computed using an updated version of PyR@TE \cite{Lyonnet:2013dna, Lyonnet:2016xiz}, 
where the scalar mixing is correctly taken into account. We compute the $\beta$ functions in the bilinear formalism.
As we show, in this  formalism, the RGEs can be put into a condensed tensor form where the $\beta$-functions 
inherit the transformation properties of the associated parameters under a change of basis.

\begingroup
\allowdisplaybreaks

\subsection{Full form}
\label{sec:RGEsFF}

\textbf{Scalar mass couplings:}
\begin{gather}
\begin{aligned}
    (16\pi^2)\,\beta(\xi_0) = {}& 24 \left(\eta_{00} \xi _0+\eta _1 \xi _1+\eta _2 \xi _2+\eta _3 \xi _3\right) \\
    +\;& 4 \xi _0 \left(E_{11}+E_{22}+E_{33}- \eta_{00} \right) - \frac{9}{2} \xi _0\left(\frac{g_1^2}{5}+g_2^2\right) \\
     +\;& \xi _0 \, \mathrm{Tr}\left[3 Y_d \left(Y_d\right){}^{\dagger }+3 \epsilon _d \left(\epsilon _d\right){}^{\dagger
   } + Y_e \left(Y_e\right){}^{\dagger } + \epsilon _e \left(\epsilon _e\right){}^{\dagger } + 3 Y_u \left(Y_u\right){}^{\dagger } + 3 \epsilon _u \left(\epsilon _u\right){}^{\dagger }\right]\\
   +\;& \xi _1 \, \mathrm{Tr}\left[3 Y_d\left(\epsilon
   _d\right){}^{\dagger } + 3\epsilon _d\left(Y_d\right){}^{\dagger } + Y_e\left(\epsilon _e\right){}^{\dagger } + \epsilon
   _e\left(Y_e\right){}^{\dagger } + 3 Y_u\left(\epsilon _u\right){}^{\dagger } + 3 \epsilon _u\left(Y_u\right){}^{\dagger }\right]\\
   +\;& i \, \xi _2 \, \mathrm{Tr}\left[ 3 \epsilon _d \left(Y_d\right){}^{\dagger }-3 Y_d \left(\epsilon _d\right){}^{\dagger } + \epsilon _e \left(Y_e\right){}^{\dagger} - Y_e \left(\epsilon _e\right){}^{\dagger } - 3 \epsilon _u \left(Y_u\right){}^{\dagger } + 3 Y_u \left(\epsilon _u\right){}^{\dagger
   }\right]\\
   +\;&\xi _3 \, \mathrm{Tr}\left[3 Y_d \left(Y_d\right){}^{\dagger } - 3 \epsilon _d \left(\epsilon _d\right){}^{\dagger } + Y_e
   \left(Y_e\right){}^{\dagger } - \epsilon _e \left(\epsilon _e\right){}^{\dagger } + 3 Y_u \left(Y_u\right){}^{\dagger } - 3 \epsilon _u
   \left(\epsilon _u\right){}^{\dagger }\right]\\
\end{aligned}\\
\begin{aligned}
    (16\pi^2)\,\beta(\xi_1) ={}& 24 \left(\eta _1 \xi _0+ E_{11} \xi _1+ E_{12} \xi _2+ E_{13} \xi _3\right) \\
    -\;& 4 \xi _1 \left(E_{11}+E_{22}+E_{33}- \eta_{00} \right)-\frac{9}{2} \xi _1\left(\frac{g_1^2}{5}+g_2^2\right) \\
    +\;& \xi _0 \, \mathrm{Tr}\left[3 Y_d\left(\epsilon
   _d\right){}^{\dagger } + 3\epsilon _d\left(Y_d\right){}^{\dagger } + Y_e\left(\epsilon _e\right){}^{\dagger } + \epsilon
   _e\left(Y_e\right){}^{\dagger } + 3 Y_u\left(\epsilon _u\right){}^{\dagger } + 3 \epsilon _u\left(Y_u\right){}^{\dagger }\right]\\
   +\;&\xi _1 \mathrm{Tr}\left[3 Y_d \left(Y_d\right){}^{\dagger }+3 \epsilon _d \left(\epsilon _d\right){}^{\dagger
   } + Y_e \left(Y_e\right){}^{\dagger } + \epsilon _e \left(\epsilon _e\right){}^{\dagger } + 3 Y_u \left(Y_u\right){}^{\dagger } + 3 \epsilon _u \left(\epsilon _u\right){}^{\dagger }\right]\\
\end{aligned}\\
\begin{aligned}
    (16\pi^2)\,\beta(\xi_2) ={}& 24 \left(\eta _2 \xi _0+ E_{12} \xi _1+ E_{22} \xi _2+ E_{23} \xi _3\right) \\
    -\;& 4 \xi _2 \left(E_{11}+E_{22}+E_{33}- \eta_{00} \right)-\frac{9}{2} \xi _2\left(\frac{g_1^2}{5}+g_2^2\right) \\
    +\;& i \, \xi _0 \, \mathrm{Tr}\left[ 3 \epsilon _d \left(Y_d\right){}^{\dagger }-3 Y_d \left(\epsilon _d\right){}^{\dagger } + \epsilon _e \left(Y_e\right){}^{\dagger} - Y_e \left(\epsilon _e\right){}^{\dagger } - 3 \epsilon _u \left(Y_u\right){}^{\dagger } + 3 Y_u \left(\epsilon _u\right){}^{\dagger
   }\right]\\
   +\;&\xi _2 \mathrm{Tr}\left[3 Y_d \left(Y_d\right){}^{\dagger }+3 \epsilon _d \left(\epsilon _d\right){}^{\dagger
   } + Y_e \left(Y_e\right){}^{\dagger } + \epsilon _e \left(\epsilon _e\right){}^{\dagger } + 3 Y_u \left(Y_u\right){}^{\dagger } + 3 \epsilon _u \left(\epsilon _u\right){}^{\dagger }\right]\\
\end{aligned}\\
\begin{aligned}
    (16\pi^2)\,\beta(\xi_3) ={}& 24 \left(\eta _3 \xi _0+ E_{13} \xi _1+ E_{23} \xi _2+ E_{33} \xi _3\right) \\
    -\;& 4 \xi _3 \left(E_{11}+E_{22}+E_{33}- \eta_{00} \right)-\frac{9}{2} \xi _3\left(\frac{g_1^2}{5}+g_2^2\right) \\
    +\;& \xi _0 \, \mathrm{Tr}\left[3 Y_d \left(Y_d\right){}^{\dagger } - 3 \epsilon _d \left(\epsilon _d\right){}^{\dagger } + Y_e
   \left(Y_e\right){}^{\dagger } - \epsilon _e \left(\epsilon _e\right){}^{\dagger } + 3 Y_u \left(Y_u\right){}^{\dagger } - 3 \epsilon _u
   \left(\epsilon _u\right){}^{\dagger }\right]\\
   +\;&\xi _3 \mathrm{Tr}\left[3 Y_d \left(Y_d\right){}^{\dagger }+3 \epsilon _d \left(\epsilon _d\right){}^{\dagger
   } + Y_e \left(Y_e\right){}^{\dagger } + \epsilon _e \left(\epsilon _e\right){}^{\dagger } + 3 Y_u \left(Y_u\right){}^{\dagger } + 3 \epsilon _u \left(\epsilon _u\right){}^{\dagger }\right]\\
\end{aligned}
\end{gather}
\textbf{Quartic couplings:}\\

For clarity, we do not display in the $\beta$-functions below the terms containing Yukawa couplings.
\begin{align}
\begin{split}
\left(16\pi^2\right)\beta(\eta_{00}) = {}& 8\, \Big(4 \eta _{00}^2+6 \left(\eta _1^2+\eta _2^2+\eta _3^2\right)+ \left(E_{11}+E_{22}+E_{33}\right) \eta _{00}\\
+\;& E_{11}^2+E_{22}^2+E_{33}^2 + 2\left(E_{12}^2+E_{13}^2 +E_{23}^2\right)\Big)\\
-\;& 9 \eta _{00} \left(\frac{g_1^2}{5}+g_2^2\right)+\frac{9}{8}   \left(\frac{3 g_1^4}{25}+g_2^4\right) 
\end{split}\raisetag{2.33\normalbaselineskip}\\
\begin{split}
\left(16\pi^2\right)\beta(\eta_1) = {}& 48 \left( \eta _{00} \eta _1 +E_{11} \eta _1 +E_{12} \eta _2+E_{13} \eta _3\right) - 9 \eta _1 \left(\frac{g_1^2}{5}+ g_2^2\right)
\end{split}\\
\begin{split}
\left(16\pi^2\right)\beta(\eta_2) = {}& 48 \left( \eta _{00} \eta _2 +E_{12} \eta _1 +E_{22} \eta _2+E_{23} \eta _3\right) - 9 \eta _2 \left(\frac{ g_1^2}{5}+ g_2^2\right)
\end{split}\\
\begin{split}
\left(16\pi^2\right)\beta(\eta_3) = {}& 48 \left( \eta _{00} \eta _3 +E_{13} \eta _1 +E_{23} \eta _2+E_{33} \eta _3\right) - 9 \eta _3 \left(\frac{ g_1^2}{5}+ g_2^2\right)
\end{split}\\
\begin{split}
\left(16\pi^2\right)\beta(E_{11}) = {} & 8 \left( 3 E_{11}^2 + 6 \eta _1^2 + 3 E_{11} \eta _{00} - E_{22} E_{11}- E_{33} E_{11}+4 E_{12}^2+4 E_{13}^2\right) \\
-\;& 9 E_{11} \left( \frac{g_1^2}{5} + g_2^2\right) + \frac{9}{20} g_1^2 g_2^2
\end{split}\\
\begin{split}
\left(16\pi^2\right)\beta(E_{22}) = {} & 8 \left( 3 E_{22}^2 + 6 \eta _2^2 + 3 E_{22} \eta _{00} - E_{11} E_{22}- E_{33} E_{22}+4 E_{12}^2+4 E_{23}^2\right) \\
-\;& 9 E_{22} \left( \frac{g_1^2}{5} + g_2^2\right) + \frac{9}{20} g_1^2 g_2^2
\end{split}\\
\begin{split}
\left(16\pi^2\right)\beta(E_{33}) = {} & 8 \left( 3 E_{33}^2 + 6 \eta _3^2 + 3 E_{33} \eta _{00} - E_{11} E_{33}- E_{22} E_{33}+4 E_{13}^2+4 E_{23}^2\right) \\
-\;& 9 E_{33} \left( \frac{g_1^2}{5} + g_2^2\right) + \frac{9}{20} g_1^2 g_2^2
\end{split}\\
\begin{split}
\left(16\pi^2\right)\beta(E_{12}) = {} & 8 \left(3 E_{12} \eta _{00}+6 \eta _1 \eta _2+3 E_{11} E_{12}+3 E_{22} E_{12}-E_{33} E_{12}+4 E_{13} E_{23}\right)\\
-\;& 9 E_{12} \left( \frac{g_1^2}{5} + g_2^2\right)
\end{split}\\
\begin{split}
\left(16\pi^2\right)\beta(E_{13}) = {} & 8 \left(3 E_{13} \eta _{00}+6 \eta _1 \eta _3+3 E_{11} E_{13}-E_{22} E_{13}+3 E_{33} E_{13}+4 E_{12} E_{23}\right)\\
-\;& 9 E_{13} \left( \frac{g_1^2}{5} + g_2^2\right)
\end{split}\\
\begin{split}
\left(16\pi^2\right)\beta(E_{23}) = {} & 8 \left(3 E_{23} \eta _{00}+6 \eta _2 \eta _3+4 E_{12} E_{13}-E_{11} E_{23}+3 E_{22} E_{23}+3 E_{23} E_{33}\right)\\
-\;& 9 E_{23} \left( \frac{g_1^2}{5} + g_2^2\right)
\end{split}
\end{align}

\subsection{Tensor form}
\label{sec:RGEsTF}

In order to express the RGEs in a compact tensor form, it is  useful to define some
abbreviations. We use in this paper the following expression for the Yukawa Lagrangian:
\begin{align}
    -\mathcal{L}_Y = \Big[ \Bar{Q}\big( Y_d\,\varphi_1 + \epsilon_d\, \varphi_2 \big) d_R + \Bar{E}\big( Y_e\,\varphi_1 + \epsilon_e\, \varphi_2 \big) e_R + \Bar{Q}\big( Y_u\,\widetilde{\varphi}_1 + \epsilon_u\, \widetilde{\varphi}_2 \big) u_R \Big] + \text{h.c.}\,,\label{YukLag}
\end{align}
where $\widetilde{\varphi}_a = \varepsilon \varphi_a^* = i \sigma_2 \varphi_a^*$ as usual.
Defining 
\begin{subequations}
    \begin{align}
        \mathcal{D}_a = \begin{pmatrix} Y_d & \epsilon_d \end{pmatrix}\ \ &, \ \ {\mathcal{D}^\dagger}^a = \begin{pmatrix} Y_d^\dagger \\ \epsilon_d^\dagger \end{pmatrix}\,,\\
        \mathcal{E}_a = \begin{pmatrix} Y_e & \epsilon_e \end{pmatrix}\ \ &, \ \ {\mathcal{E}^\dagger}^a = \begin{pmatrix} Y_e^\dagger \\ \epsilon_e^\dagger \end{pmatrix}\,,\\
        \mathcal{U}^a = \begin{pmatrix} Y_u \\ \epsilon_u \end{pmatrix}\ \ &, \ \ \mathcal{U}^\dagger_a = \begin{pmatrix} Y_u^\dagger & \epsilon_u^\dagger \end{pmatrix}\,,
    \end{align}
\end{subequations}
allows one to put the Lagrangian \eqref{YukLag} in the following form :
\begin{align}
    -\mathcal{L}_Y = \Big[ \Bar{Q}\;\mathcal{D}_a\,\varphi^a\; d_R + \Bar{E}\;\mathcal{E}_a\,\varphi^a\;e_R + \Bar{Q}\;\mathcal{U}^a\,\widetilde{\varphi}_a\; u_R \Big] + \text{h.c}\,.
\end{align}
We note that $\mathcal{U}$ has an upper index, since $\widetilde{\varphi}$ transforms in the anti-fundamental representation of the global $U(2)$ symmetry.

We can now define the following tensors, which will be useful to express the $\beta$-functions in a concise way (where $N_c = 3$): 
\begin{subequations}\label{tTrace}
\begin{equation}
    \tensor{\mathcal{T}}{_a^b} = N_c \,\mathrm{Tr}\big[\mathcal{U}^\dagger_a \, \mathcal{U}^b\big] + 
    N_c\, \mathrm{Tr}\big[\mathcal{D}^{\dagger b} \mathcal{D}_a\big] + \mathrm{Tr}\big[\mathcal{E}^{\dagger b} \mathcal{E}_a\big]\, ,
\end{equation}
\begin{equation}
    \tensor{\mathcal{T}}{_a^b_c^d} = N_c\, \mathrm{Tr}\big[\mathcal{U}_a^\dagger \, \mathcal{U}^b\, \mathcal{U}_c^\dagger \, \mathcal{U}^d \big] +
    N_c \, \mathrm{Tr}\big[\mathcal{D}^{\dagger b} \mathcal{D}_a \mathcal{D}^{\dagger d}  \mathcal{D}_c \big] + \mathrm{Tr}\big[\mathcal{E}^{\dagger b} \mathcal{E}_a\, \mathcal{E}^{\dagger d} \mathcal{E}_c \big]\, ,
\end{equation}
\begin{equation}
    \mathcal{T}_{\mathcal{U}\mathcal{D}} = N_c\, \mathrm{Tr}\big[\mathcal{U}^\dagger_a \mathcal{D}_b \,( \mathcal{D}^{\dagger a}\mathcal{U}^b - \mathcal{D}^{\dagger b}\mathcal{U}^a)\big] = N_c\, \varepsilon^{a b}\varepsilon_{c d}\, \mathrm{Tr}\big[ \mathcal{U}^\dagger_a \mathcal{D}_b \mathcal{D}^{\dagger c}  \mathcal{U}^d \big]\, .
\end{equation}
\end{subequations}\\
\textbf{Scalar mass couplings:}

\begin{align}
        (16\pi^2)\,\beta(\xi_0) &= 24 (\eta_{00}\xi_0 + \eta_i \xi_i) + 4 \xi_0 \big( \mathrm{Tr}(E) - \eta_{00}\big) - \frac{9}{2} \xi _0\left(\frac{g_1^2}{5}+g_2^2\right) 
        \nonumber \\
        &\phantom{=} +\big(\xi_0\, \tensor{\delta}{^a_b} + \xi_i\tensor{(\sigma_i)}{^a_b}\big)\tensor{\mathcal{T}}{_a^b}\ , \\
        (16\pi^2)\,\beta(\xi_i) &= 24 (\eta_{i}\xi_0 + E_{ij} \xi_j) - 4 \xi_i \big( \mathrm{Tr}(E) - \eta_{00}\big)- \frac{9}{2} \xi _i \left(\frac{g_1^2}{5}+g_2^2\right) 
        \nonumber\\
        &\phantom{=} + \big(\xi_i\,\tensor{\delta}{^a_b} + \xi_0 \tensor{(\sigma_i)}{^a_b}\big)\tensor{\mathcal{T}}{_a^b}\, .
\end{align}
\textbf{Quartic couplings:}

\begin{align}
\begin{split}
    (16\pi^2)\,\beta(\eta_{00}) = {}& 32 \eta _{00}^2 + 48 \eta_k \eta^k + 8  \eta _{00}\, \mathrm{Tr}(E)  + 8 \mathrm{Tr}(E^2) 
    \\
    & -9 \eta _{00} \left(\frac{g_1^2}{5}+g_2^2\right)+\frac{9}{8}   \left(\frac{3 g_1^4}{25}+g_2^4\right)
    \\
    & + 2\big( \eta_{00}\, \tensor{\delta}{^a_b} + \eta_{k} \tensor{(\sigma_k)}{^a_b} \big)\,\tensor{\mathcal{T}}{_a^b} - \frac{1}{2}\big[\tensor{(\sigma_0)}{^a_b}\tensor{(\sigma_0)}{^c_d}\big]\, \tensor{\mathcal{T}}{_a^b_c^d} + \mathcal{T}_{\mathcal{U}\mathcal{D}}\, ,
\end{split}\\[.25cm]
\begin{split}
    (16\pi^2)\,\beta(\eta_{i}) = {}& 48\eta_{00}\eta_i + 48 \tensor{E}{_i_j}\eta_j - 9\eta_i\left(\frac{g_1^2}{5}+g_2^2\right)\\
    & + \big(2\eta_i \tensor{\delta}{^a_b} + \eta_{00} \tensor{(\sigma_i)}{^a_b} + E_{ij} \tensor{(\sigma_j)}{^a_b} \big)\,\tensor{\mathcal{T}}{_a^b} - \frac{1}{2}\big[\tensor{(\sigma_0)}{^a_b}\tensor{(\sigma_i)}{^c_d}\big]\, \tensor{\mathcal{T}}{_a^b_c^d}\, ,
\end{split}\\[.25cm]
\begin{split}
    (16\pi^2)\,\beta(\tensor{E}{_i_j}) = {}& 32 \tensor{E}{_i_k} \tensor{E}{_k_j} - 8 \mathrm{Tr}(E) \tensor{E}{_i_j} + 24 \eta_{00} \tensor{E}{_i_j} + 48 \eta_i \eta_j 
    \\
    & - 9 \tensor{E}{_i_j}\left(\frac{g_1^2}{5}+g_2^2\right) + \tensor{\delta}{_i_j} \left( \frac{9}{20} g_1^2 g_2^2 \right) 
    \\
    & + \big( 2 E_{ij} \tensor{\delta}{^a_b} + \eta_{i} \tensor{(\sigma_j)}{^a_b} + \eta_{j} \tensor{(\sigma_i)}{^a_b} \big)\,\tensor{\mathcal{T}}{_a^b} - \frac{1}{2}\big[\tensor{(\sigma_i)}{^a_b}\tensor{(\sigma_j)}{^c_d}\big]\, \tensor{\mathcal{T}}{_a^b_c^d} - \delta_{ij} \mathcal{T}_{\mathcal{U}\mathcal{D}}\, .
\end{split}
\end{align}
\textbf{Yukawa couplings:}\\

For completeness, we include here the $\beta$-function of the up-type Yukawa couplings in tensor form:
\begin{align}
\begin{split}
	(16\pi^2)\,\beta\left(\mathcal{U}^a\right) = {}&  \mathcal{U}^a\,\mathcal{U}^\dagger_b \mathcal{U}^b +  \mathcal{D}_b \mathcal{D}^{\dagger b}\,\mathcal{U}^a -\left(\frac{17}{20}g_1^2 + \frac{9}{4}g_2^2 + 8 g_3^2 \right)\mathcal{U}^a \\
	+\;&\Big[\tensor{\delta}{^a_b} \tensor{\delta}{^c_d} +\tensor{(\sigma_i)}{^a_b} \tensor{(\sigma^i)}{^c_d}\Big]\, \left(\frac{1}{4} \mathcal{U}^b\,\mathcal{U}^\dagger_c \mathcal{U}^d - \frac{3}{2} \mathcal{D}_c \mathcal{D}^{\dagger d}\,\mathcal{U}^b + \frac{1}{2} \tensor{\mathcal{T}}{_c^d}\,\mathcal{U}^b \right)\, .
\end{split}
\end{align}

\section{The top Yukawa coupling}
\label{top}

Let us compute the top-quark mass in the THDM type III.
With the help of the expression of the Higgs doublets in the Yukawa Lagrangian~\eqref{YukLag} we have
\begin{equation}
    - \mathcal{L}_Y = \Bar{Q}_L \big( y' \widetilde{\varphi}_1' + \epsilon' \widetilde{\varphi}'_2\big) t_R + \mathrm{h.c.} \, ,
\end{equation}
where primed quantities are computed in the Higgs basis. In the unitary gauge at the vacuum, this relation becomes    
\begin{equation}
       - \langle \mathcal{L}_Y \rangle
    = \frac{v_0}{\sqrt{2}} y'_t \, \Bar{t}_L t_R  + \mathrm{h.c.}\, .
\end{equation}
We see that, as in the Standard Model, the top mass is given by 
\begin{equation}
    m_t^2 = \frac{v_0^2}{2} |y_t'|^2 \ .
\end{equation}

%
%

We now want to show how a basis transformation of the Higgs-boson doublets \eqref{EWbasisChange}
affects the top-quark mass.
Under a change of basis, the top Yukawa couplings transforms as
\begin{align}
    Y \equiv \begin{pmatrix} y_t \\ \epsilon_t \end{pmatrix} \rightarrow Y' = \begin{pmatrix} y'_t \\ \epsilon'_t \end{pmatrix} = U Y \, .
\end{align}
Expressed in terms of the original parameters, the top mass squared is therefore
\begin{align}
    m_t^2 &= \frac{v_0^2}{2} |y_t'|^2 \nonumber \\
    &= \frac{v_0^2}{2} \left|y_t\, \cos{\beta} + \epsilon_t\, \sin{\beta}\, e^{-i\zeta}\right|^2\nonumber \\
    &= \frac{v_0^2}{2} \Big( |y_t|^2\cos^2\beta + 2 \Re\big[y_t\epsilon^*_t \, e^{i\zeta}\big]\cos{\beta}\sin{\beta} + |\epsilon_t|^2 \sin^2\beta \Big) \, .
    \label{topMass}
\end{align}

This general form can be specialized to other types of THDM by imposing that either $y_t$ or $\epsilon_t$ vanishes in the original basis. In this case $\beta$ can be understood as the usual physical parameter related to the ratio of the vevs.

\clearpage

\endgroup



\bibliographystyle{JHEP}
\bibliography{MPP2}

\providecommand{\href}[2]{#2}\begingroup\raggedright\begin{thebibliography}{10}

\bibitem{Bennett:1993pj}
D.~L. Bennett and H.~B. Nielsen, \emph{{Predictions for nonAbelian fine
  structure constants from multicriticality}},
  \href{https://doi.org/10.1142/S0217751X94002090}{\emph{Int. J. Mod. Phys.}
  {\bfseries A9} (1994) 5155}
  [\href{https://arxiv.org/abs/hep-ph/9311321}{{\ttfamily hep-ph/9311321}}].

\bibitem{Bennett:1996hx}
D.~L. Bennett, \emph{{Multiple point criticality, nonlocality, and fine tuning
  in fundamental physics: Predictions for gauge coupling constants gives
  $\alpha^{-1} = 136.8 \pm 9$}}, Ph.D. thesis, Bohr Inst., 1996.
\newblock \href{https://arxiv.org/abs/hep-ph/9607341}{{\ttfamily
  hep-ph/9607341}}.

\bibitem{Bennett:1996vy}
D.~L. Bennett and H.~B. Nielsen, \emph{{Gauge couplings calculated from
  multiple point criticality yield $\alpha^{-1} = 136.8 \pm 9$: At last the
  elusive case of U(1)}},
  \href{https://doi.org/10.1142/S0217751X9900155X}{\emph{Int. J. Mod. Phys.}
  {\bfseries A14} (1999) 3313}
  [\href{https://arxiv.org/abs/hep-ph/9607278}{{\ttfamily hep-ph/9607278}}].

\bibitem{Bennett:2003yr}
D.~Bennett and H.~Nielsen, \emph{{The multiple point principle: Realized vacuum
  in nature is maximally degenerate}}, {\emph{Bled Workshops Phys.} {\bfseries
  4} (2003) 235}.

\bibitem{Froggatt:1995rt}
C.~D. Froggatt and H.~B. Nielsen, \emph{{Standard model criticality prediction:
  Top mass $173 \pm 5$ GeV and Higgs mass $135 \pm 9$ GeV}},
  \href{https://doi.org/10.1016/0370-2693(95)01480-2}{\emph{Phys. Lett.}
  {\bfseries B368} (1996) 96}
  [\href{https://arxiv.org/abs/hep-ph/9511371}{{\ttfamily hep-ph/9511371}}].

\bibitem{Degrassi:2012ry}
G.~Degrassi, S.~Di~Vita, J.~Elias-Miro, J.~R. Espinosa, G.~F. Giudice,
  G.~Isidori et~al., \emph{{Higgs mass and vacuum stability in the Standard
  Model at NNLO}}, \href{https://doi.org/10.1007/JHEP08(2012)098}{\emph{JHEP}
  {\bfseries 08} (2012) 098} [\href{https://arxiv.org/abs/1205.6497}{{\ttfamily
  1205.6497}}].

\bibitem{Tanabashi:2018oca}
{\scshape Particle Data Group} collaboration, \emph{{Review of Particle
  Physics}}, \href{https://doi.org/10.1103/PhysRevD.98.030001}{\emph{Phys.
  Rev.} {\bfseries D98} (2018) 030001}.

\bibitem{Chatrchyan:2012xdj}
{\scshape CMS} collaboration, \emph{{Observation of a New Boson at a Mass of
  125 GeV with the CMS Experiment at the LHC}},
  \href{https://doi.org/10.1016/j.physletb.2012.08.021}{\emph{Phys. Lett.}
  {\bfseries B716} (2012) 30}
  [\href{https://arxiv.org/abs/1207.7235}{{\ttfamily 1207.7235}}].

\bibitem{Aad:2012tfa}
{\scshape ATLAS} collaboration, \emph{{Observation of a new particle in the
  search for the Standard Model Higgs boson with the ATLAS detector at the
  LHC}}, \href{https://doi.org/10.1016/j.physletb.2012.08.020}{\emph{Phys.
  Lett.} {\bfseries B716} (2012) 1}
  [\href{https://arxiv.org/abs/1207.7214}{{\ttfamily 1207.7214}}].

\bibitem{Lee:1973iz}
T.~D. Lee, \emph{{A Theory of Spontaneous T Violation}},
  \href{https://doi.org/10.1103/PhysRevD.8.1226}{\emph{Phys. Rev.} {\bfseries
  D8} (1973) 1226}.

\bibitem{Bernreuther:1998rx}
W.~Bernreuther and O.~Nachtmann, \emph{{Flavor dynamics with general scalar
  fields}}, \href{https://doi.org/10.1007/s100520050535,
  10.1007/s100529900029}{\emph{Eur. Phys. J.} {\bfseries C9} (1999) 319}
  [\href{https://arxiv.org/abs/hep-ph/9812259}{{\ttfamily hep-ph/9812259}}].

\bibitem{Branco:2011iw}
G.~C. Branco, P.~M. Ferreira, L.~Lavoura, M.~N. Rebelo, M.~Sher and J.~P.
  Silva, \emph{{Theory and phenomenology of two-Higgs-doublet models}},
  \href{https://doi.org/10.1016/j.physrep.2012.02.002}{\emph{Phys. Rept.}
  {\bfseries 516} (2012) 1} [\href{https://arxiv.org/abs/1106.0034}{{\ttfamily
  1106.0034}}].

\bibitem{Gunion:1989we}
J.~F. Gunion, H.~E. Haber, G.~L. Kane and S.~Dawson, \emph{{The Higgs Hunter's
  Guide}}, {\emph{Front. Phys.} {\bfseries 80} (2000) 1}.

\bibitem{Froggatt:2004st}
C.~D. Froggatt, L.~V. Laperashvili, R.~B. Nevzorov, H.~B. Nielsen and M.~Sher,
  \emph{{The Two Higgs doublet model and the multiple point principle}},
  {\emph{Bled Workshops Phys.} {\bfseries 5} (2004) 28}
  [\href{https://arxiv.org/abs/hep-ph/0412333}{{\ttfamily hep-ph/0412333}}].

\bibitem{McDowall:2018ulq}
J.~McDowall and D.~J. Miller, \emph{{High Scale Boundary Conditions in Models
  with Two Higgs Doublets}},
  \href{https://doi.org/10.1103/PhysRevD.100.015018}{\emph{Phys. Rev.}
  {\bfseries D100} (2019) 015018}
  [\href{https://arxiv.org/abs/1810.04518}{{\ttfamily 1810.04518}}].

\bibitem{Nagel:2004sw}
F.~Nagel, \emph{{New aspects of gauge-boson couplings and the Higgs sector}},
  Ph.D. thesis, Heidelberg U., 2004.

\bibitem{Maniatis:2006fs}
M.~Maniatis, A.~von Manteuffel, O.~Nachtmann and F.~Nagel, \emph{{Stability and
  symmetry breaking in the general two-Higgs-doublet model}},
  \href{https://doi.org/10.1140/epjc/s10052-006-0016-6}{\emph{Eur. Phys. J.}
  {\bfseries C48} (2006) 805}
  [\href{https://arxiv.org/abs/hep-ph/0605184}{{\ttfamily hep-ph/0605184}}].

\bibitem{Nishi:2006tg}
C.~C. Nishi, \emph{{CP violation conditions in N-Higgs-doublet potentials}},
  \href{https://doi.org/10.1103/PhysRevD.76.119901,
  10.1103/PhysRevD.74.036003}{\emph{Phys. Rev.} {\bfseries D74} (2006) 036003}
  [\href{https://arxiv.org/abs/hep-ph/0605153}{{\ttfamily hep-ph/0605153}}].

\bibitem{Ma:2009ax}
E.~Ma and M.~Maniatis, \emph{{Symbiotic Symmetries of the Two-Higgs-Doublet
  Model}}, \href{https://doi.org/10.1016/j.physletb.2009.11.056}{\emph{Phys.
  Lett.} {\bfseries B683} (2010) 33}
  [\href{https://arxiv.org/abs/0909.2855}{{\ttfamily 0909.2855}}].

\bibitem{Froggatt:2008am}
C.~D. Froggatt, R.~Nevzorov, H.~B. Nielsen and D.~Thompson, \emph{{On the
  origin of approximate custodial symmetry in the Two-Higgs Doublet Model}},
  \href{https://doi.org/10.1142/S0217751X09047442}{\emph{Int. J. Mod. Phys.}
  {\bfseries A24} (2009) 5587}
  [\href{https://arxiv.org/abs/0806.3190}{{\ttfamily 0806.3190}}].

\bibitem{Maniatis:2007vn}
M.~Maniatis, A.~von Manteuffel and O.~Nachtmann, \emph{{CP violation in the
  general two-Higgs-doublet model: A Geometric view}},
  \href{https://doi.org/10.1140/epjc/s10052-008-0712-5}{\emph{Eur. Phys. J.}
  {\bfseries C57} (2008) 719}
  [\href{https://arxiv.org/abs/0707.3344}{{\ttfamily 0707.3344}}].

\bibitem{Froggatt:2006zc}
C.~D. Froggatt, L.~Laperashvili, R.~Nevzorov, H.~B. Nielsen and M.~Sher,
  \emph{{Implementation of the multiple point principle in the two-Higgs
  doublet model of type II}},
  \href{https://doi.org/10.1103/PhysRevD.73.095005}{\emph{Phys. Rev.}
  {\bfseries D73} (2006) 095005}
  [\href{https://arxiv.org/abs/hep-ph/0602054}{{\ttfamily hep-ph/0602054}}].

\bibitem{Lyonnet:2013dna}
F.~Lyonnet, I.~Schienbein, F.~Staub and A.~Wingerter, \emph{{PyR@TE:
  Renormalization Group Equations for General Gauge Theories}},
  \href{https://doi.org/10.1016/j.cpc.2013.12.002}{\emph{Comput. Phys. Commun.}
  {\bfseries 185} (2014) 1130}
  [\href{https://arxiv.org/abs/1309.7030}{{\ttfamily 1309.7030}}].

\bibitem{Lyonnet:2016xiz}
F.~Lyonnet and I.~Schienbein, \emph{{PyR@TE 2: A Python tool for computing RGEs
  at two-loop}}, \href{https://doi.org/10.1016/j.cpc.2016.12.003}{\emph{Comput.
  Phys. Commun.} {\bfseries 213} (2017) 181}
  [\href{https://arxiv.org/abs/1608.07274}{{\ttfamily 1608.07274}}].

\end{thebibliography}\endgroup

\end{document}